\documentclass[lineno]{jfm}
\usepackage{pdflscape} 
\usepackage{booktabs} 
\usepackage{longtable}     
\usepackage[table]{xcolor}  
\usepackage{colortbl}       
\usepackage{array}         
\usepackage{graphicx}
\usepackage{newtxtext}
\usepackage{newtxmath}
\usepackage{natbib}
\usepackage[dvipsnames]{xcolor}
\usepackage{fdsymbol}
\usepackage{pifont}
\usepackage{hyperref}
\usepackage{url}
\usepackage{float}
\usepackage{soul}
\usepackage{tikz}   
\usepackage{xcolor} 
\newcommand{\RomanNumeralCaps}[1]
\linenumbers

\newcommand{\re}{Re_{\tau}}
\newcommand{\zp}{z^{+}}
\newcommand{\up}{U^{+}}
\newcommand{\ut}{U_{\tau}}
\newcommand{\zc}{z^{+}_{c}}

\definecolor{OliveGreen}{RGB}{0,200,0}

\definecolor{or}{rgb}{1.0, 0.5, 0.0}
\definecolor{mg}{rgb}{0.9, 0.0, 0.9}
\definecolor{yy}{rgb}{1.0, 1.0, 0.0}

\definecolor{k1}{rgb}{0.6, 0.6, 0.6}
\definecolor{k2}{rgb}{0.3, 0.3, 0.3}
\definecolor{k3}{rgb}{0.0, 0.0, 0.0}

\definecolor{r1}{rgb}{1 ,0.6, 0.6}
\definecolor{r2}{rgb}{1, 0.3, 0.3}
\definecolor{r3}{rgb}{1, 0.0, 0.0}

\definecolor{g1}{rgb}{0.6, 0.84, 0.6}
\definecolor{g2}{rgb}{0.3, 0.72, 0.3}
\definecolor{g3}{rgb}{0.0, 0.60, 0.0}

\hypersetup{
    colorlinks = true,
    urlcolor   = blue,
    citecolor  = black,
}

\title{High-Reynolds-number turbulent boundary layers under adverse pressure gradients. 
Part 2. A composite mean velocity profile}
\author{Ahmad Zarei, Mitchell Lozier, Rahul Deshpande \and Ivan Marusic
\corresp{\email{imarusic@unimelb.edu.au}}}
\affiliation{Department of Mechanical Engineering, The University of Melbourne, Victoria 3010, Australia}

\begin{document}
\maketitle

\begin{abstract}
A robust composite mean velocity profile is developed for turbulent boundary layers (TBLs) subjected to adverse pressure gradients (APGs), extending the composite formulation for generic pressure-gradient TBLs proposed by \citeauthor{nickels} (\textit{J.\ Fluid Mech.}, vol.\ 521, 2004). Several modifications are introduced to capture characteristic features of APG flows.
A new parameter accounts for pressure-gradient history effects in the wake region, a velocity-overshoot function is incorporated in the inner region, and the wake function is reformulated using an independent, physically motivated definition of the boundary-layer thickness.
A compilation of APG TBL datasets from the literature, including the new dataset presented in Part~1, is used to assess and refine the formulation.
The resulting composite profile contains three physically meaningful parameters that capture pressure-gradient effects on the mean velocity profile, determined through nonlinear curve fitting.
These parameters form the basis of a framework for identifying `well-behaved' APG TBLs and as well as quantifying the strength of pressure-gradient history effects.
The composite profile further enables reliable estimation of the friction velocity and boundary-layer thickness in well-behaved APG TBLs, providing a practical tool for scaling analyses when these quantities are not directly measurable.
Its analytical form yields improved estimates of mean velocity gradients, facilitating evaluation of the indicator function and identification of inflection points.
Finally, the formulation provides predictive capability for both the coefficients and spatial extent of the logarithmic region of the mean streamwise velocity profile, enabling evaluation of its universality in high-Reynolds-number APG TBLs. This reveals that the von Kármán coefficient approaches an invariant value of $\kappa \approx 0.39$ at sufficiently high Reynolds numbers, independent of pressure-gradient effects.
\end{abstract}

\section{Introduction}

In Part~1 of this study, the influence of local and upstream pressure-gradients (PG) on high-Reynolds-number adverse-pressure-gradient (APG) turbulent boundary layers (TBLs) was examined, demonstrating that PG history can significantly modify the mean velocity profile.
Building on those findings, the present paper focuses on the wall-normal structure of the mean streamwise velocity in APG TBLs and the development of a composite representation of this profile.

The mean velocity profile is of particular interest because it is directly associated with quantities of engineering relevance, such as the skin-friction drag generated by the TBL \citep{panton2005}.
More broadly, a well-resolved mean velocity profile is a fundamental component in the development of mixing-length and eddy-viscosity models, which can be used to predict the Reynolds stress distribution of the TBL \citep{subrahmanyam2022integral}.
It also plays a central role in resolvent-based descriptions of wall-bounded turbulence, where the mean profile acts as the base flow that governs the amplification and organisation of coherent motions \citep{sharma2013coherent}.

These uses make accurate knowledge of the mean velocity profile valuable for both simulations and experiments, particularly when quantities such as the Reynolds stresses or the wall-shear stress cannot be measured directly.
The importance of such approaches increases with the friction Reynolds number ($\re=\delta U_{\tau}/\nu$), as numerical simulations become more expensive and experimental measurements of Reynolds stresses become increasingly difficult.
Here, $\delta$ denotes the boundary-layer thickness \citep[determined in this manuscript using the method of][]{lozier2025defining}, $U_{\tau}$ is the friction velocity (related to the wall-shear stress through $\tau_w=\rho U_\tau^2$), $\rho$ is the density, and $\nu$ is the kinematic viscosity.

For these reasons, the development of analytical expressions describing the mean velocity profile has been an enduring objective in wall-bounded turbulence research \citep{panton2005}.
Such descriptions take a variety of forms \citep[e.g.\ the integral formulations of][]{subrahmanyam2022integral, cantwell2022, yang2024, shu2025mean}; one widely used approach is the composite formulation.
In composite models, the mean profile is constructed by combining expressions that represent the dominant physics of different regions of the boundary layer (e.g.\ the inner, overlap, and wake regions) into a single continuous profile \citep[see][]{nickels}.
These formulations are typically expressed in terms of normalised variables, with the mean streamwise velocity $U$ and wall-normal coordinate $z$ written in inner units (denoted by the superscript `$+$'), scaled by the viscous velocity ($U_\tau$) and length ($\nu/U_\tau$) scales.

It is well known that TBL statistics, including the mean velocity profile, depend on several flow parameters such as $\re$, streamwise pressure gradients and surface roughness.
Consequently, developing a robust and general composite representation of the mean velocity profile is challenging.
Such a formulation must capture the underlying physics across the different regions of the boundary layer while remaining consistent with high-fidelity datasets spanning a broad range of flow conditions.

\subsection{Composite mean velocity profile formulations for the TBL}

We briefly review the development of composite formulations for the mean velocity profile in turbulent boundary layers. 
In the classical view, the composite profile follows a wall--wake structure consisting of two functional relationships commonly referred to as the `law of the wall' and the `law of the wake' \citep{coles1956law, marusic2010wall}:
\begin{equation}
    \up =f(\zp) + \mathrm{g}(\Pi,\eta),
    \label{eq:coles}
\end{equation}
where the wall function, $f$, depends on the viscous-scaled wall-normal distance ($\zp=z\ut/\nu$), while the wake function, $\mathrm{g}$, depends on the outer-scaled wall-normal distance ($\eta=z/\delta$) and Coles' wake-strength parameter, $\Pi$.

The wall function can be further decomposed into two fundamental scaling laws obtained through a range of arguments \citep[see][]{millikan1938, clauser1956turbulent, coles1956law, panton2005, marusic2010wall}. 
The first applies very near the wall ($\zp \lesssim 5$, i.e.\ in the viscous sublayer), where
\begin{equation}
    \up=\zp,
    \label{eq:closeWall}
\end{equation}
as required in the limit $\zp \rightarrow 0$. 
The second applies farther from the wall while $\eta$ remains small:
\begin{equation}
    \up = \frac{1}{\kappa} \,\ln ( \zp ) + B,
    \label{eq:loglaw}
\end{equation}
where $\kappa$ and $B$ are traditionally referred to as the von K\'arm\'an and additive constants \citep{marusic2013logarithmic}, respectively, although recent studies suggest that they should be regarded as coefficients that may depend on flow and boundary conditions \citep{Monkewitz_Nagib_2023}.
This classical logarithmic law is observed in the overlap region of canonical high-$\re$ TBLs and has been studied extensively \citep{marusic2010wall}.

Between the viscous sublayer and overlap region lies the buffer region, where neither scaling law applies exclusively. 
Its functional form is not well defined physically, and some studies \citep{mesolayer} have further identified `mesolayers' between the buffer and overlap regions. 
Accordingly, many studies have sought to combine the two classical scaling laws into a single expression for the near-wall mean velocity. 
For example, \citet{van1956Profile} derived a shear-stress relation that, when integrated, yields an integral expression for the streamwise velocity, although its practical use is not straightforward. 
Alternative expressions, such as those of \citet{spalding1961single} and \citet{musker1979explicit}, provide a single representation for the viscous sublayer and buffer regions (collectively termed the `inner' region) and asymptotically recover the logarithmic law with increasing wall-normal distance.

Following these ideas, many composite formulations take the form
\begin{equation}
    \up = \up_{\text{inner}}(\zp)+\up_{\text{overlap}}(\zp,\eta)+\up_{\text{wake}}(\eta,\Pi),
    \label{eq:TotProf}
\end{equation}
where separate analytical expressions describe the inner, overlap and wake regions of the TBL \citep[see][]{chauhan2009criteria}. 
If the boundary-layer thickness and friction velocity are known, these formulations contain only one free parameter, $\Pi$, which may be obtained by fitting the composite profile to the measured mean velocity profile. 
Such formulations are therefore referred to here as \textit{one-parameter composite profiles}. 
They typically provide good fits for `well-behaved' canonical TBLs \citep[i.e.\ smooth-wall, zero-pressure-gradient (ZPG) flows with minimal history effects; see][]{chauhan2009criteria}, although some limitations remain.

For example, the boundary-layer thickness used in these formulations does not always correspond to a commonly or easily measured quantity (e.g.\ $\delta_{99}$), and is sometimes treated as an additional fitting parameter. 
Further uncertainty arises in the wake region. 
The original wake function, $\mathcal{W}$, introduced by \citet{coles1956law} was sinusoidal, such that the wake contribution was written as
\begin{equation} 
    \up_{\text{wake}}= \frac{2\Pi}{\kappa} \mathcal{W}(\eta).
    \label{eq:colesWake}
\end{equation}
Subsequent studies have instead used exponential functions \citep{chauhan2007composite, nickels}, polynomial functions \citep{jones2001sink}, and other physically motivated models \citep[for example, based on turbulent/non-turbulent interface (TNTI) interactions;][]{krug2017}. 
Although these forms can produce good fits for particular TBLs, ambiguity regarding a generally applicable wake function persists.

For pressure-gradient (PG) TBLs specifically, the mean velocity profile depends not only on Reynolds number but also on both the local streamwise PG \citep{nagano, deshpande, harun, drozdz2020description, monty2011} and the upstream PG history \citep{bobke2017history, pozuelo, romero202history, vinuesa2017revisiting}. 
However, there is currently no established mathematical framework for explicitly describing the mean velocity profile of generic PG TBLs. 
Although one-parameter composite profiles were not intended only for canonical flows, they generally do not describe non-canonical TBLs well, including certain regimes of streamwise PG and/or surface roughness. 
For example, in PG flows without a wake region (e.g.\ sink flows), the overlap expression in \eqref{eq:TotProf} grows unboundedly, whereas the mean velocity should approach a constant value at the boundary-layer edge (i.e.\ $\up \rightarrow \up_\infty$ as $\eta \rightarrow 1$). 
These limitations have motivated new composite formulations for specific non-canonical TBL classes, including generic PG TBLs \citep{nickels}.

Accounting for streamwise PGs in a composite mean velocity formulation requires at least one additional free parameter beyond $\Pi$, yielding a \textit{two-parameter composite profile}. 
Here the local PG is quantified using Clauser's pressure-gradient parameter,
\[
\beta = \frac{\delta^\ast}{\tau_w}\left(\frac{dP}{dx}\right),
\]
where $\delta^\ast$ denotes the displacement thickness and $dP/dx$ the streamwise gradient of the static pressure. 
The local PG strength may also be expressed using viscous scaling alone:
\begin{equation} 
    p_{x}^{+} =  \frac{\nu}{\rho u_\tau^3} \left( \frac{dP}{dx} \right).
    \label{eq:px}
\end{equation}
This viscous-scaled PG parameter, $p_{x}^{+}$, is important because it arises naturally in composite formulations, specifically in the inner-region expression, through a Taylor-series expansion of the boundary-layer equations near the wall \citep{nickels}. 
The PG-related terms in this expansion are often neglected because $p_x^{+}$ is small in high-$\re$ TBLs ($p_{x}^{+}=\beta/\delta^{*+}$) and/or in flows with low-to-moderate PGs. 
However, these terms become significant in low-$\re$ PG TBLs and in flows subjected to strong PGs \citep{nickels}.

\begin{figure}
    \captionsetup{width=1.00\linewidth}
    \begin{center}
    \includegraphics[width=0.85\textwidth]{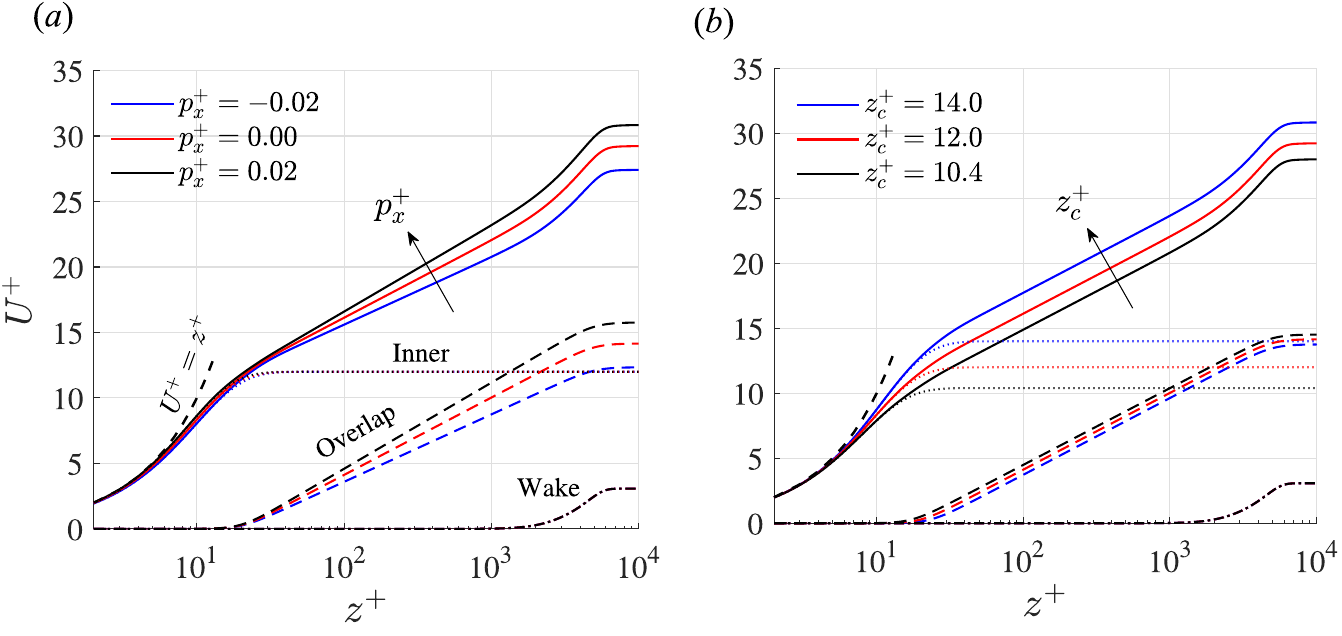}
    \end{center}
    \caption{Contributions of the inner, overlap and wake expressions from \citet{nickels} as a function of (\textit{a}) $p_x^{+}$ and (\textit{b}) $\zc$. The composite mean velocity profile (solid lines) is the sum of the three individual expressions. A baseline dataset with $\re=5000$ was used to generate these profiles.}
    \label{fig1}
\end{figure}

Accordingly, \citet{nickels} reintroduced the $p_x^{+}$-dependent terms and introduced a new parameter, $\zc=z_c U_{\tau}/\nu$, into the inner and overlap expressions of the composite profile:
\begin{equation}
    \up = \up_{\text{inner}} (\zp,p_x^+,\zc)+\up_{\text{overlap}} (\zp,p_x^+,\zc,\eta)+\up_{\text{wake}}(\eta,\Pi),
    \label{eq:TotPressureG}
\end{equation}
The parameter $\zc$, referred to as the `sublayer thickness', accounts for PG-induced changes to the inner and overlap regions of the mean velocity profile. 
The details of this formulation are given by \citet{nickels}; here, only the key features relevant to the present study are summarised. 
The contributions of the inner, overlap, and wake terms in Nickels' formulation are shown in figure~\ref{fig1}, together with the effects of varying $\zc$ and $p_x^{+}$. 
The shape of the inner-region profile depends on the magnitude of $p_x^+$ through the inclusion of the PG-related terms from the Taylor expansion (figure~\ref{fig1}a), while the inner-region profile asymptotes to a constant value equal to the sublayer thickness, $\zc$ (figure~\ref{fig1}b).

A fundamental relationship between $p_x^+$ and $\zc$ can be obtained by following the arguments of \citet{clauser1954turbulent} and \citet{van1963boundary}, which introduce a critical Reynolds number,
\[
R_c=z_c U_{\tau,c}/\nu,
\]
where $z_c$ is a critical wall-normal location at which the sublayer becomes unstable, and $U_{\tau,c}=\sqrt{\tau(z=z_c)/\rho}$. 
Within Nickels' framework, the total shear stress near the wall may be written as
\begin{equation}
    \frac{\tau}{\tau_w} = -\overline{uw}^+ + \frac{\partial U^+}{\partial z^+} \;\approx\; 1 + p_x^+ z^+,
    \label{eq:shear}
\end{equation}
and substitution into the critical-Reynolds-number relation yields
\begin{equation}
    p_x^+ z_c^{+3} + z_c^{+2} - R_c^2 = 0.
    \label{eq:pxNickles}
\end{equation}
Here, $R_c$ is assumed to be universal for wall-bounded flows, with a value of approximately $12$.

Nickels also showed that in the high-Reynolds-number limit $\zc$ is directly related to the classical additive constant $B$ of the logarithmic law, indicating that $\zc$ plays an important role in describing deviations from classical logarithmic scaling in PG TBLs. 
The formulation further prescribes changes in the von K\'arm\'an coefficient proportional to $p_x^+$, suggesting that at sufficiently high Reynolds numbers the von K\'arm\'an coefficient remains quasi-invariant and asymptotes to a constant value, $\kappa$. 
In addition, the overlap expression in \eqref{eq:TotPressureG} does not grow unboundedly (figure~\ref{fig1}), thereby avoiding the non-physical behaviour of some composite profiles, particularly in flows with a weak or absent wake region such as sink flows.

These predictions were tested experimentally in Part~1, where the universality of $\kappa$ and the dependence of the additive constant, $B$, on $\beta$ were investigated using a methodology that imposed moderate APG conditions with minimal PG-history effects at high $\re$. 
The results showed that, at high $\re$, $\kappa$ remained invariant within experimental uncertainty, while $B$ decreased systematically with both local PG strength and PG history, consistent with the general framework of \citet{nickels}.

In summary, Nickels' formulation may be categorised as a \textit{two-parameter composite profile}, with free parameters $\zc$ and $\Pi$ accounting for variations in streamwise PG and Reynolds number. 
In practice, however, as with one-parameter composite profiles, $\delta$ is often treated as an additional fitting parameter. 
Moreover, although Nickels' formulation was developed for generic PG TBLs, it does not explicitly account for PG history, and it does not adequately describe the mean velocity profile of high-Reynolds-number PG TBLs in all cases, since $p_x^+$ is a \emph{local} and \emph{viscous-scaled} parameter, as demonstrated further in §\ref{sec:NickelsLimits}. 
For example, in the wake region, PG history was found in Part~1 to modify both the vertical displacement of the mean velocity, through changes in $\Pi$ \citep[see also][]{preskett2025effects}, and the horizontal stretching of the wake profile, through changes in its $\eta$ dependence. 
Similarly, in the overlap region, PG history introduces additional deviations from the $\beta$-based trend observed for the additive constant in Part~1. 
These observations suggest that at least one additional parameter is required to describe a broader range of PG TBLs, particularly those at high $\re$ or influenced by substantial upstream PG histories.

Motivated by the framework of \citet{nickels} and the findings of Part~1, the growing database of APG TBLs in the literature provides a foundation for developing a more robust composite profile for generic APG TBLs.
Past experimental and numerical investigations \citep[e.g.][]{nagano, monty2011, bobke2017history, pozuelo, vila2017, perry_streamwise_2002, preskett2025effects, zarei2026_part1} have provided new insight into the behaviour of the mean velocity profile across a wide range of Reynolds numbers and APG conditions.
Building on these datasets and physical insights, the present study develops a composite mean velocity profile formulation applicable across a broad range of Reynolds numbers and APG conditions while accounting for PG history effects with minimal additional complexity.
Analysis of the available datasets indicates that a two-parameter composite profile is insufficient to describe the range of PG TBL behaviour considered here.
Accordingly, the formulation of \citet{nickels} is extended by introducing one additional parameter required to capture these effects, yielding a \textit{three-parameter composite profile}.

The resulting formulation also provides several practical utilities.
In particular, it enables estimation of $\ut$ and $\delta$ for APG TBL datasets in which these quantities are difficult to measure experimentally, especially for high $Re_{\tau}$ TBLs or when PG histories are minimal or well controlled.
The analytical form of the profile further allows evaluation of quantities involving wall-normal derivatives of the mean velocity, such as the indicator function, which are often difficult to obtain from experimental data.
Finally, the additional parameter introduced to account for PG history effects can be interpreted in terms of an `effective' local pressure gradient, providing new physical insight into the development of APG TBLs.

The remainder of the paper is organised as follows.
Section~2 introduces the datasets used to assess and refine the composite profile formulation.
Section~3 examines the limitations of existing composite profiles and motivates the modifications introduced here.
Section~4 presents the new composite profile formulation and the associated fitting procedure.
Section~5 evaluates the accuracy of the formulation and interprets trends in the fitted parameters across the compiled datasets.
Section~6 discusses practical utilities of the composite profile for analysing APG TBL datasets.

\section{Datasets}

To assess the robustness of the proposed composite profile across a broad range of flow conditions, a carefully selected compilation of APG TBL datasets from the literature is considered.
These datasets span more than three decades of experimental and numerical investigations and include well-resolved large-eddy simulations (LES) and direct-numerical simulations (DNS), together with experimental measurements, with and without PG history effects, covering a broad range of $\beta$ and $\re$, as listed in table~\ref{tab:database}.
Additional datasets (e.g. \citealp{marusic2015, orlu2013, li2021experimental}) are also considered to highlight the influence of varying tripping or surface roughness conditions, which are distinct from PG-related influences.
Considering APG TBLs both with and without pronounced PG history effects is a key aspect of the database assembled here, enabling the influence of upstream PG history to be decoupled from the effects of local flow conditions for parameters of the composite profile, such as $\zc$.
This compilation of datasets, together with the composite profile framework described below, also provides a useful means of contextualising the effects of different PG histories across studies reported in the literature.

\begin{table}
    \centering 
    \renewcommand{\arraystretch}{1.0} 
    \setlength{\tabcolsep}{5pt}       
    \small 
    \begin{tabular}{cccccc}
    \hline
    Reference & Database & $\beta$ & $\re$ & Cases & Symbol \\
    \hline
    \citet{maruvsic1995wall}    & EXP   & 0-5       & 2840-4170  & 6  & \tikz\draw[draw=black,fill=yy] (-0.10,-0.11) -- (0.10,-0.11) -- (0,0.11) -- cycle;\\
    \citet{nagano}              & EXP   & 0.77-5.32 & 420-600    & 3  &  \tikz\draw[draw=red,fill=red] (-0.10,-0.11) -- (0.10,-0.11) -- (0,0.11) -- cycle;\\
    \citet{monty2011}           & EXP   & 0-1.52    & 2170-2970  & 2  & \tikz\draw[draw=g3,fill=g3] (-0.10,-0.11) -- (0.10,-0.11) -- (0,0.11) -- cycle;\\
     \citet{orlu2013}           & EXP   & 0         & 1150-1970  & 3  & \tikz\draw[draw=mg,fill=mg] (-0.10,-0.11) -- (0.10,-0.11) -- (0,0.11) -- cycle;\\
    \citet{eitel2014simulation} & LES   & 0         & 1200-3140  & 4  & \tikz\draw[draw=or,fill=or] (-0.10,-0.11) -- (0.10,-0.11) -- (0,0.11) -- cycle;\\
    \citet{marusic2015}         & EXP   & 0         & 2820-13500 & 12 &\tikz\draw[draw=blue,fill=blue] (-0.10,-0.11) -- (0.10,-0.11) -- (0,0.11) -- cycle;\\
    \citet{bobke2017history}    & LES   & 1.03-1.64 & 490-770    & 2  & \tikz\draw[draw=black, fill=yy] (0,0.12) -- (0.09,0) -- (0,-0.12) -- (-0.09,0) -- cycle;\\
    \citet{yoon}                & DNS   & 1.45      & 830        & 1  & \tikz\draw[draw=red, fill=red] (0,0.12) -- (0.09,0) -- (0,-0.12) -- (-0.09,0) -- cycle;\\
    \citet{vila2020}            & EXP   & 0-2.19    & 1430-5070  & 9  & \tikz\draw[draw=g3, fill=g3] (0,0.12) -- (0.09,0) -- (0,-0.12) -- (-0.09,0) -- cycle;\\
    \citet{volino2020non}       & EXP   & 0-6.6     & 1060-1450  & 8  & \tikz\draw[draw=mg, fill=mg] (0,0.12) -- (0.09,0) -- (0,-0.12) -- (-0.09,0) -- cycle;\\
    \citet{li2021experimental}  & EXP   & 0         & 8610-23930 & 8  & \tikz\draw[draw=red, fill=red] (-0.10,0.11) -- (0.10,0.11) -- (0,-0.11) -- cycle;\\
    \citet{pozuelo}             & LES   & 1.03-1.65 & 490-770    & 2  & \tikz\draw[draw=or, fill=or] (0,0.12) -- (0.09,0) -- (0,-0.12) -- (-0.09,0) -- cycle;\\
    \citet{romero2022}          & EXP   & 0.9-1.77  & 7100-7770  & 5  & \tikz\draw[draw=blue, fill=blue] (0,0.12) -- (0.09,0) -- (0,-0.12) -- (-0.09,0) -- cycle;\\
    \citet{gungor2024}          & DNS   & 0-0.2     & 860-1520   & 2  & \tikz\draw[draw=black, fill=yy] (-0.10,0.11) -- (0.10,0.11) -- (0,-0.11) -- cycle;\\
    \citet{preskett2025effects} & EXP   & 0         & 6840-9440  & 6  & \tikz\draw[draw=or, fill=or] (-0.10,0.11) -- (0.10,0.11) -- (0,-0.11) -- cycle;\\
    \citet{zarei2026_part1} & EXP   & 0-1.47    & 4000-10680 & 23 &  See Part 1\\
    \hline
    \end{tabular}
    \caption{Summary of flow conditions and symbols for the datasets considered in this study.}
    \label{tab:database}
\end{table}

\section{Formulation of a new composite profile for APG TBLs} 
\label{sec:NickelsLimits}
\begin{figure}
    \captionsetup{width=1.00\linewidth}
    \begin{center}
    \includegraphics[width=1.00\textwidth]{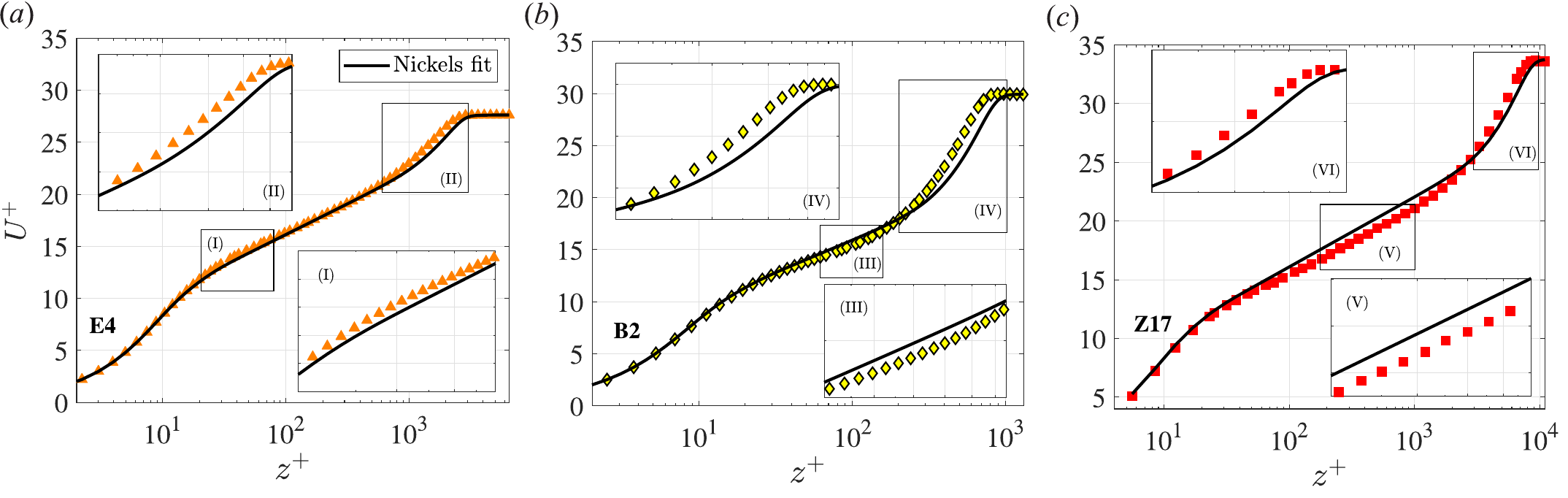}
    \end{center}
    \caption{Mean velocity data and corresponding composite profile \eqref{eq:TotPressureG} for (\textit{a}) a well-resolved LES ZPG dataset, (\textit{b}) an APG dataset influenced by history effects, and (\textit{c}) a high-$\re$ APG dataset with minimal PG history.}
    \label{fig4}
\end{figure}

To motivate key modifications to existing composite formulations for PG TBLs, the profile of \citet{nickels} is applied to several representative datasets from table~\ref{tab:database}.
This assessment examines its accuracy and limitations across different flow conditions and across the various regions of the TBL.
Specifically, the formulation of \citet{nickels} is fitted to high-resolution ZPG numerical data from \citet{eitel2014simulation}, high-$\re$ APG experimental data from \citet{zarei2026_part1}, and APG LES data with PG history effects from \citet{bobke2017history}, as shown in figure~\ref{fig4}.
The two-character designations for each profile correspond to specific cases listed in table~\ref{tab:database}, with nomenclature and specific details of individual cases documented in Appendix~\ref{apx:B}.

Starting with the well-resolved ZPG dataset of \citet{eitel2014simulation} (figure~\ref{fig4}a), the composite profile accurately captures the mean velocity in the overlap region and near the wall ($\zp<10$).
However, visible discrepancies remain in the buffer (inset I) and wake (inset II) regions.
The discrepancy in the buffer region arises from a velocity overshoot relative to the overlap-region scaling (i.e.\ the classical logarithmic law), which has been observed in both experimental and numerical studies \citep{monkewitz2007self} but is not captured by the inner-region expression in \eqref{eq:TotPressureG}.

In the wake region, discrepancies arise when the boundary-layer thickness ($\delta$) is not treated as a fitted parameter.
In the present study, $\delta$ is instead defined using the physically motivated approach derived from the skewness profile of the mean streamwise velocity fluctuations, and turbulent/non-turbulent interface framework of \citet{lozier2025defining}.
Applying this independently determined $\delta$ within the formulation of \citet{nickels} reveals that the wake function in \eqref{eq:TotPressureG} is not well suited to adopt this definition, as demonstrated consistently in figures~\ref{fig4}(a--c).
\citet{nickels} also noted this limitation; allowing $\delta$ to vary as a fitted parameter introduced at least a 3\% difference relative to the $\delta_{98.5}$ definition, which may become larger when alternative definitions of $\delta$ are used.

For the APG datasets, additional discrepancies appear in the overlap region for cases with strong PG history effects (inset III, figure~\ref{fig4}b) or at high Reynolds number (inset V, figure~\ref{fig4}c).
These deviations arise from the overlap formulation of \citet{nickels}, which relates shifts in the mean velocity to $p_x^+$.
Because $p_x^+$ diminishes with increasing $\re$ (irrespective of the magnitude of $dP/dx$) and reflects only the local PG, it does not capture the cumulative influence of PG history.

The discrepancy in the wake region is particularly pronounced for the case with strong PG history effects (inset IV, figure~\ref{fig4}b).
This behaviour is consistent with the findings of Part~1, which showed that PG history significantly modifies both the magnitude and the distribution of the mean velocity profile in the wake region.

\subsection{The new composite profile formulation}

The observations described above motivate modifications to the formulation of \citet{nickels}.
Specifically, the present formulation introduces a velocity overshoot in the inner region, reformulates the wake function to be consistent with the physically motivated definition of $\delta$ adopted here, and incorporates recent physical insights regarding pressure-gradient effects in the overlap region (at high $\re$) and PG-history effects in the wake region.

Accordingly, we propose a modified composite profile expressed as
\begin{equation}
    \up = \up_{\text{inner}} (\zp,p_x^+,\zc)+\up_{\text{overlap}} (\zp,p_x^+,\zc,\eta,C_{H_w})+\up_{\text{wake}}(\eta,C_{H_w},\Pi).
    \label{eq:PresentFit}
\end{equation}

The specific modifications to each component of the formulation are described below.

\subsubsection{Addition of a wake-stretching parameter}
An additional parameter, $C_{H_w}$, is introduced as a `wake-stretching' parameter to account for PG-history effects observed in the wake region.
The coefficients in the wake function are also revised to ensure consistency with the boundary-layer thickness definition of \citet{lozier2025defining}.
Accordingly, the outer-scaled wall-normal coordinate $\eta$ is rescaled within the wake-region expression ($\up_{\text{wake}}$), resulting in an effective coordinate, 
\[ 
\eta_{\rm eff} = 1.455 C_{H_w} \eta. \]

\subsubsection{Addition of an overshoot function}

To account for the velocity overshoot observed in the buffer region, a Gaussian-type function is introduced following the approach of \citet{monkewitz2007self} and \citet{chauhan2009criteria}, who incorporated a similar term into the inner-region expression of \citet{musker1979explicit}.
The overshoot function is written as
\begin{equation}
    f(\zc) \exp \left( \frac{-\ln^2(z^+/38)}{0.835} \right),
    \label{eq:f}
\end{equation}
where $f$ represents the maximum overshoot relative to the baseline profile of \citet{musker1979explicit}.
For ZPG flow, $f \approx 0.2$, consistent with the findings of \citet{chauhan2009criteria}.
However, this value depends on the pressure gradient (i.e.\ varies with $\zc$), as demonstrated in appendix~\ref{apx:overshoot}.

The contribution of the overshoot function is restricted to the inner region, $9 \leq \zp \leq 150$, with its maximum occurring at $\zp = 38$.
This ensures that the inner-region expression asymptotes to $\up=\zc$ with increasing $z$ (figure~\ref{fig5}), consistent with the framework of \citet{nickels} and maintaining compatibility with the overlap-region formulation.

\subsection{Linking local and upstream PGs to $\zc$}

As demonstrated in Part~1 and by \citet{bobke2017history}, PG-history effects can produce differences in the overlap-region mean velocity profiles of APG TBLs even when the local flow conditions ($\re$ and $\beta$) are matched.
These differences appear primarily as variations in the additive constant $B$ \citep[or equivalently $\zc$, see][]{nickels}.
To capture this behaviour, we introduce the concept of an effective pressure-gradient parameter,
\[ \beta_{\rm eff} = \beta C_{H_i}, \]
which combines the local PG parameter, $\beta$, with a PG-history parameter, $C_{H_i}$, to represent the combined influence of local and upstream pressure gradients on the inner and overlap regions.

Importantly, $C_{H_i}$ is not introduced as an additional free parameter of the composite profile.
Rather, it is obtained from the fitting process and provides a means of linking local and upstream PG effects to $\zc$, thereby offering physical insight into the flow conditions experienced by the TBL.
To model this relationship, we follow the framework of \citet{nickels}, as described by \eqref{eq:shear} and \eqref{eq:pxNickles}.
While $p_x^+$ plays an important role in the near-wall shear-stress expression for APG TBLs \eqref{eq:shear} (particularly for $z^+<5$), the resulting model for the sublayer thickness \eqref{eq:pxNickles} does not capture the changes in $\zc$ (or equivalently $B$) observed for high-$\re$ TBLs with moderate APGs in Part~1.

Accordingly, we propose that $\zc$ is more appropriately described as a function of $\beta_{\rm eff}$, consistent with the results of Part~1 and previous studies.
This leads to a modified form of \eqref{eq:pxNickles}:
\begin{equation}
    c\, \beta_{\rm eff}\, z_c^{+3} + z_c^{+2} - R_c^2 = 0,
    \label{eq:model}
\end{equation}
whose root predicts $\zc$ for a given $\beta_{\rm eff}$.
This model yields excellent agreement with experimental data for both low- and high-$\re$ APG TBLs when $c=0.0145$, as demonstrated by the dashed line in figure~\ref{fig9}(a).

The history parameter $C_{H_i}$ can then be expressed in terms of $\zc$ and $\beta$ using \eqref{eq:model}:
\begin{equation}
    \begin{array}{ll}
    C_{H_i} = \dfrac{144 - z_c^{+2}}{c \beta z_c^{+3}} & \text{for } \beta \neq 0 \\[10pt]
    C_{H_i} = \dfrac{12}{z_c^{+}} & \text{for } \beta = 0
    \end{array}
    \label{eq:Chi}
\end{equation}

To avoid the singularity that arises for nominally ZPG cases ($\beta=0$), the second expression compares the fitted $\zc$ directly with the canonical ZPG value $\zc=12$.
For APG TBLs with minimal upstream PG-history effects, both history parameters introduced here should be approximately unity ($C_{H_i} \approx 1$ and $C_{H_w} \approx 1$).
Deviations from unity therefore indicate a significant influence of PG history on the TBL.

Notably, $C_{H_i}$ and $C_{H_w}$ are independent parameters that can be interpreted in terms of the characteristic time scales associated with different PG effects.
Specifically, $C_{H_i}$ reflects the relatively rapid response of the inner region to changes in PG conditions, whereas $C_{H_w}$ represents the slower adjustment of the wake region, consistent with the observations reported in Part~1.

Incorporating these modifications, the proposed composite profile formulation becomes
\begin{multline}
    U_{\text{inner}}^+ = \zc \Bigg[ 1 - \Bigg( 1 + 2 \left( \frac{z^+}{\zc} \right) 
    + \frac{1}{2} \left( 3 - p_x^+ \zc \right) \left( \frac{z^+}{\zc} \right)^2  
    - \frac{3}{2} p_x^+ \zc \left( \frac{z^+}{\zc} \right)^3 \Bigg) 
    e^{- \frac{3z^+}{\zc}} \Bigg]  \\
    + \left(0.23 + 0.5 \tanh\left(31 \, \frac{144 - {\zc}^2}{{\zc}^3}\right)\right) \exp \left( \frac{-\ln^2(z^+/38)}{0.835} \right)
    \label{eq:PresentInner}
\end{multline} 
\begin{equation}
    U^+_{\text{overlap}} = \frac{\sqrt{1+p_x^+\zc}} {6\kappa} \ln \left( \frac {1+(0.6(z^+/\zc))^6} {1 + \eta_{\rm{eff}}^6} \right); 
    \label{eq:Presentoverlap}
\end{equation}
\begin{equation}
    U^+_{\text{wake}} = \frac{2\Pi}{\kappa} \left( 1 - \exp\left({\frac{-15(\eta_{\rm{eff}}^4 + \eta_{\rm{eff}}^8)}{1 + 15\eta_{\rm{eff}}^3}} \right) \right).
    \label{eq:Presentwake}
\end{equation}

\begin{figure}
    \captionsetup{width=1.00\linewidth}
    \begin{center}
    \includegraphics[width=1.00\textwidth]{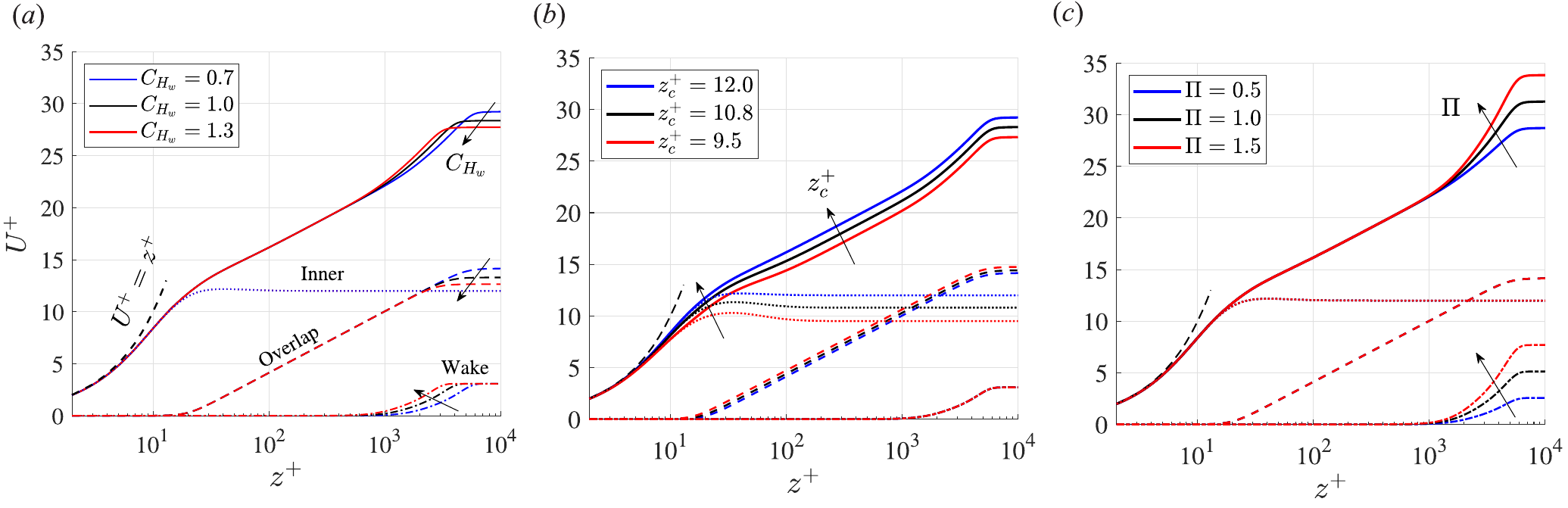}
    \end{center}
    \caption{Contributions of the inner \eqref{eq:PresentInner}, overlap \eqref{eq:Presentoverlap} and wake \eqref{eq:Presentwake} expressions to the present composite profile as a function of (\textit{a}) the wake history factor $C_{H_w}$, (\textit{b}) $\zc$, and (\textit{c}) $\Pi$. A reference dataset with $\re=5000$ was used to generate these profiles.}
    \label{fig5}
\end{figure}

Figure~\ref{fig5} illustrates the influence of the three free parameters ($C_{H_w}$, $\zc$, and $\Pi$) on the resulting composite mean velocity profile.
Variations in $C_{H_w}$ and $\Pi$ primarily reshape the wake region, influencing both the extent and magnitude of the wake.
In contrast, variations in $\zc$ modify the overlap region (through shifts in the mean velocity) and the inner region (through the magnitude of the buffer-region overshoot and the near-wall profile shape).

\subsection{Note for high-Reynolds-number TBLs}

In the high-$\re$ limit, the proposed composite profile formulation simplifies, particularly in the overlap region.
Specifically, as $\re \rightarrow \infty$, $p_x^+ \rightarrow 0$ and $\eta_{\rm eff} \rightarrow 0$, while $\zp \rightarrow \infty$ (since $\delta^+ \rightarrow \infty$).
In this limit, the inner-region expression asymptotes to $\up=\zc$ for $\zp \gg \zc$.
Consequently, the overlap-region expression \eqref{eq:Presentoverlap} reduces to
\begin{equation}
    U^+ = \frac{1}{\kappa} \ln(z^+) - \underbrace{\frac{1}{\kappa} \ln(\zc/0.6) + \zc}_{B},
    \label{eg:overlapHigh}
\end{equation}
which has the form of the classical logarithmic law.
Here $\kappa$ remains constant, while the additive constant $B$ depends only on $\zc$ (i.e.\ the pressure gradient), consistent with the results of Part~1.
For ZPG TBLs, when $\zc=12$, \eqref{eg:overlapHigh} reduces to the classical log-law with $\kappa=0.39$ and $B=4.3$, consistent with \citet{marusic2013logarithmic}.

\subsection{Fitting method}
\label{secFit}

We now outline practical considerations for applying the proposed composite mean velocity profile to APG TBL datasets.
The primary inputs may be expressed in normalised form as $U^+$, $z^+$, $\re$ and $p_x^+$ (or equivalently in dimensional form as $U$, $z$, $U_\tau$, $\delta$, $dP/dx$, $\nu$ and $\rho$).
The three parameters optimised when fitting the composite profile to a dataset are the sublayer thickness $\zc$ (which also determines $\beta_{\rm eff}$ and $C_{H_i}$), together with the wake parameters $\Pi$ and $C_{H_w}$.

For all cases considered here, the parameters were obtained using a nonlinear least-squares curve-fitting procedure applied to the measured mean velocity profiles.
To facilitate application of this method, an example implementation is provided as a \href{https://cocalc.com/share/public_paths/cd3b9c28f1e14b1de432482bc8009c461a69cfad/Fitting_CompleteParams.ipynb}{notebook}.
Normalised mean velocity data, together with the corresponding flow parameters described above, can be supplied as inputs.
The notebook then performs the nonlinear fit to determine the optimal values of the three free parameters in the composite profile formulation \eqref{eq:PresentFit} and plots the resulting fitted profile.

The fitting procedure was tested for all datasets considered in this study (table~\ref{tab:Alldatasests}) to confirm that the optimisation remained robust across a wide range of flow conditions.
In addition, for TBLs with minimal or well-controlled PG history effects (i.e.\ $C_{H_i}=C_{H_w}=1$), the composite profile may alternatively be used to estimate $U_\tau$ and $\delta$, as discussed in §\ref{sec:utilities}.
An implementation of this procedure is also provided in a separate \href{https://cocalc.com/share/public_paths/cd3b9c28f1e14b1de432482bc8009c461a69cfad/Fitting_UnknownParams.ipynb}{notebook}, where dimensional mean velocity data can be supplied when $U_\tau$ and $\delta$ are not known a priori.

\section{Quality and interpretation of the new composite profile}
\label{secQuality}

In this section, the accuracy of the new composite profile is evaluated using a range of ZPG and APG datasets, with and without PG history effects, and trends in the optimised fitting parameters are examined. Appendix \ref{apx:B} provides a complete list of parameters obtained from the fitting procedure for the cases considered.

\begin{figure}
    \captionsetup{width=1.00\linewidth}
    \begin{center}
    \includegraphics[width=0.80\textwidth]{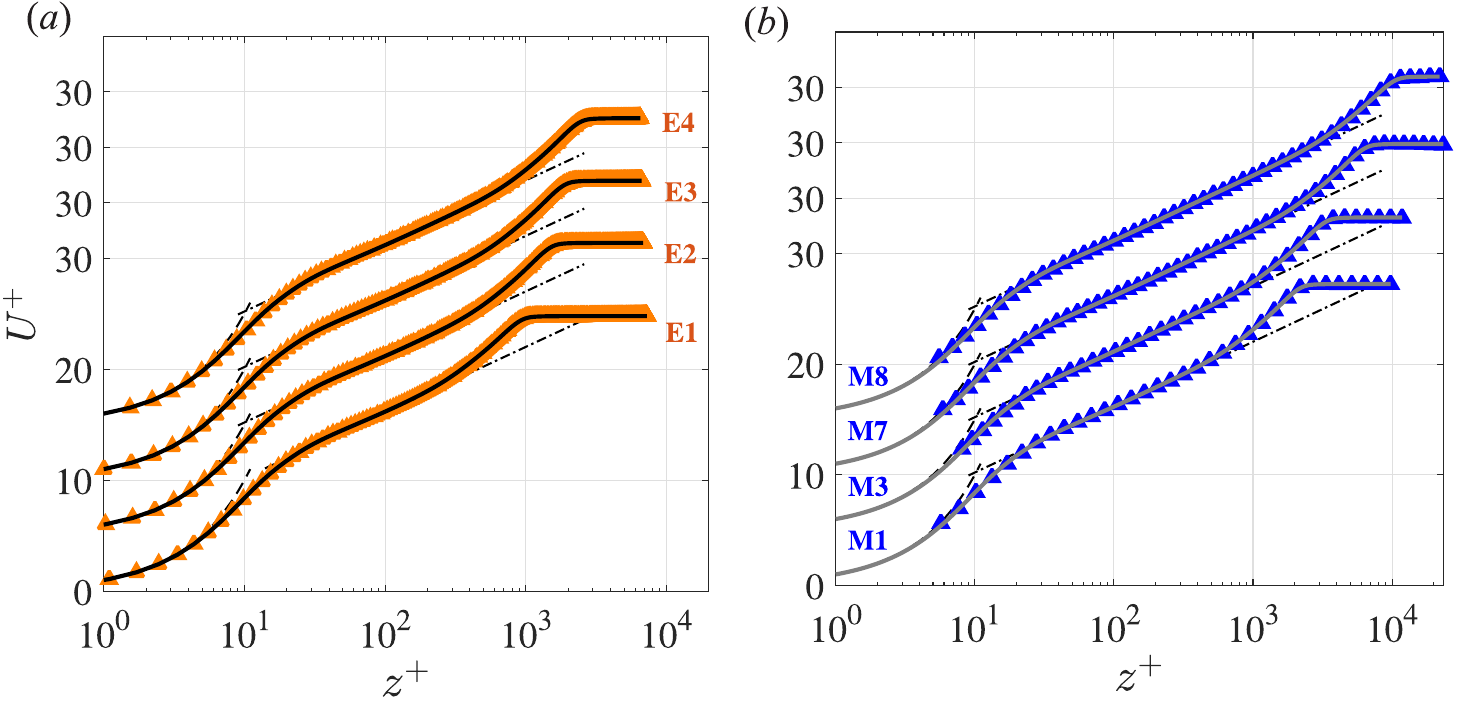}
    \end{center}
    \caption{Validation of the new composite profile for ZPG TBLs with minimal PG history from (\textit{a}) the low-$\re$ LES dataset of \citet{eitel2014simulation}, and (\textit{b}) the high-$\re$ dataset of \citet{marusic2015}. Individual profiles are offset by five units in the vertical direction. Dashed lines represent \eqref{eq:closeWall} and dash-dotted lines represent \eqref{eq:loglaw}.}
    \label{fig6}
\end{figure}

The composite profile is first fitted to a series of ZPG datasets with well-controlled PG histories, as shown in figure~\ref{fig6}.
The optimised profiles match the mean velocity data well across all cases and regions of the TBL.
The resulting fitting parameters confirm that $\zc=12$ for ZPG TBLs, independent of data resolution and database type (i.e.\ experimental or numerical), consistent with \citet{nickels}.
Moreover, the resulting history parameters are found to be $C_{H_w} \approx C_{H_i} \approx 1$ for all cases, confirming that PG history effects are minimal. 

\begin{figure}
    \captionsetup{width=1.00\linewidth}
    \begin{center}
    \includegraphics[width=0.90\textwidth]{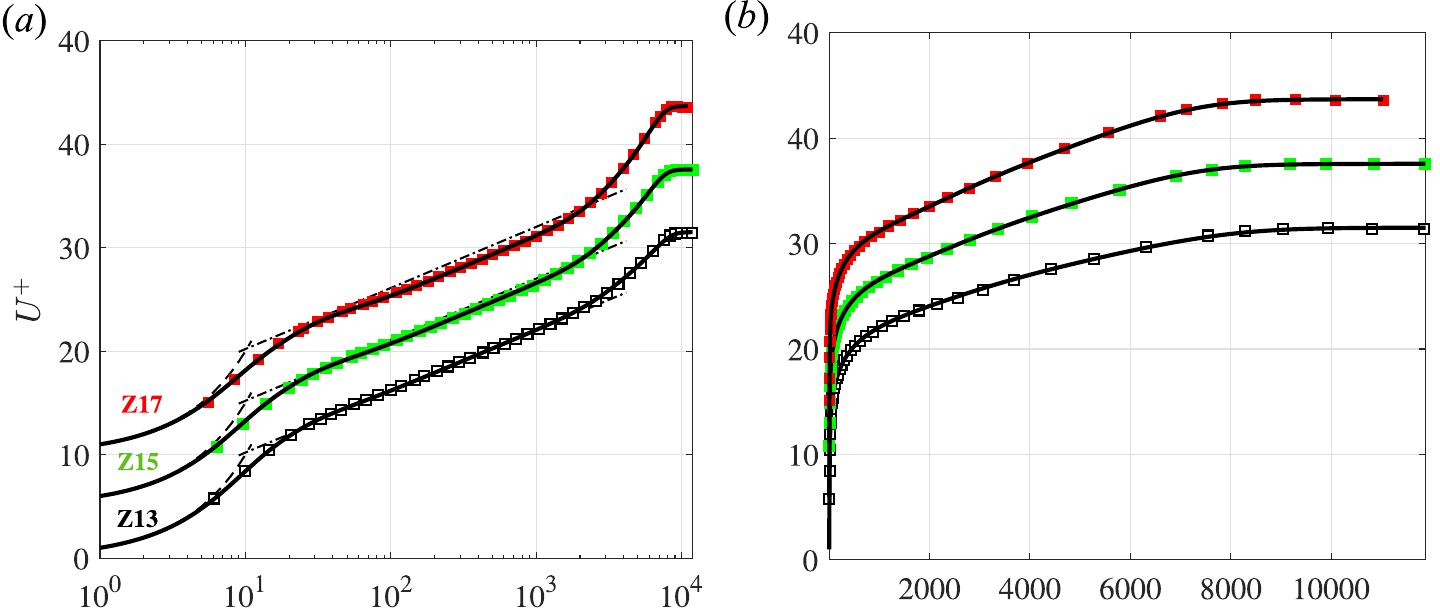}
    \end{center}
    \caption{Validation of the new composite profile for high-$\re$ moderate-APG TBLs with minimal PG history effects from Part 1. The mean velocity profiles are plotted in both (\textit{a}) logarithmic and (\textit{b}) linear scaling to confirm the quality of the composite profile across all regions of the TBL. Dashed lines represent \eqref{eq:closeWall} and dash-dotted lines represent \eqref{eq:loglaw}.}
    \label{fig7}
\end{figure}

Next, the composite profile is fitted to the high-$\re$ moderate-APG dataset with minimal PG history effects from Part~1, as shown in figure~\ref{fig7}.
These cases are of particular interest because they enable PG-related trends in the optimised fitting parameters to be examined without the additional influence of PG history effects.
The profiles are presented using both linear and logarithmic scaling of the wall-normal distance ($\zp$) to demonstrate the quality of the fit across all regions of the TBL.
As with the ZPG datasets above, the resulting fitting parameters indicate that $C_{H_w} \approx 1$ for these APG datasets, confirming minimal effects of PG history.

\begin{figure}
    \captionsetup{width=1.00\linewidth}
    \begin{center}
    \includegraphics[width=0.90\textwidth]{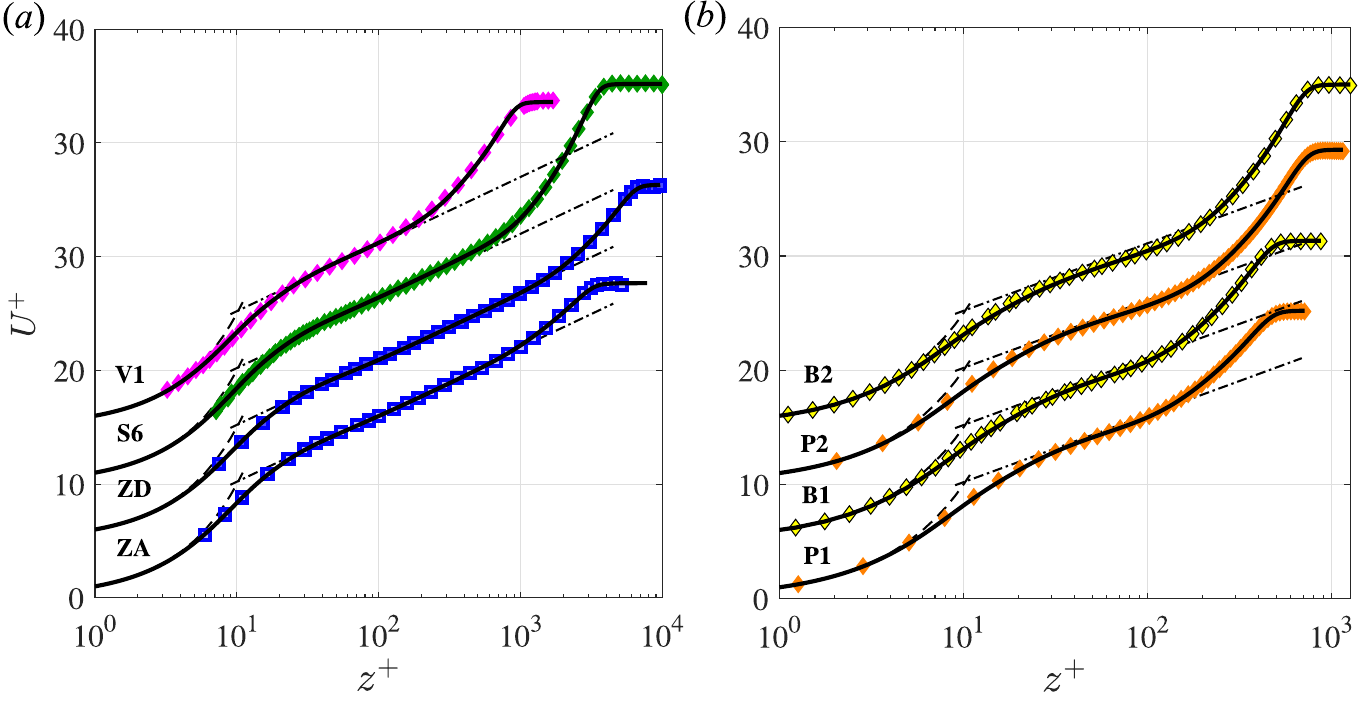}
    \end{center}
    \caption{Validation of the new composite profile for a range of (\textit{a}) experimental and (\textit{b}) numerical APG TBL datasets with PG history effects.  Dashed lines represent \eqref{eq:closeWall} and dash-dotted lines represent \eqref{eq:loglaw}.}
    \label{fig7a}
\end{figure}

\begin{figure}
    \captionsetup{width=1.00\linewidth}
    \begin{center}
    \includegraphics[width=0.45\textwidth]{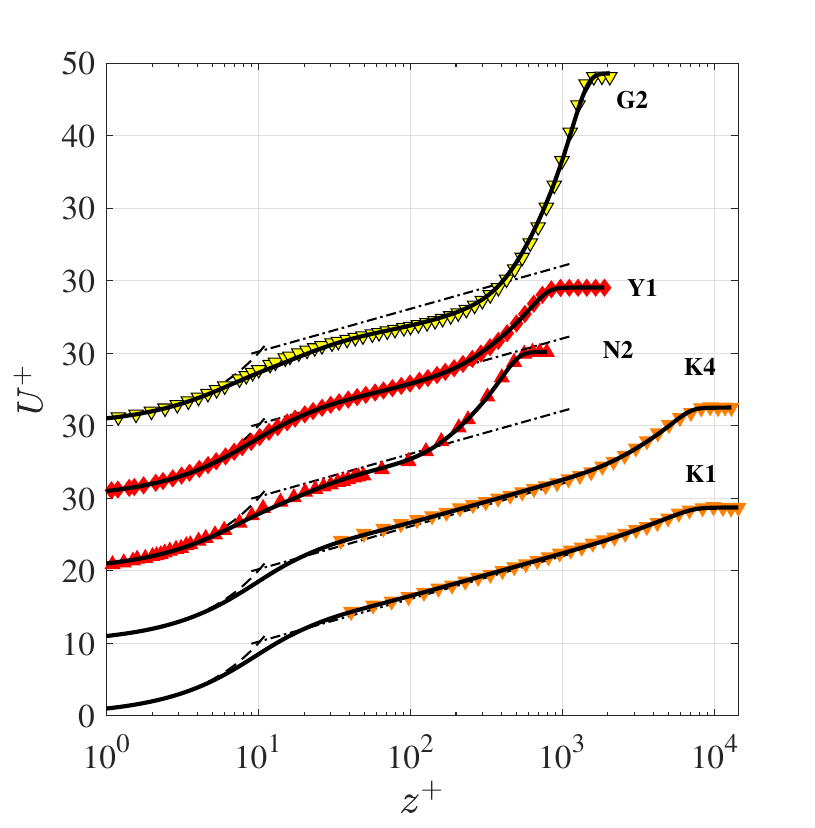}
    \end{center}
    \caption{Validation of the new composite profile for additional APG TBL datasets where $\delta$ is not known. Dashed lines represent \eqref{eq:closeWall} and dash-dotted lines represent \eqref{eq:loglaw}.}
    \label{fig14}
\end{figure}

Finally, the composite profile is fitted to select APG datasets with varying PG history effects that span a broad range of $\re$ and $\beta$, as shown in figures~\ref{fig7a} and \ref{fig14}.
For clarity, the experimental and numerical datasets are plotted separately in figure~\ref{fig7a}.
The fitted composite profiles match the mean velocity data well in each case, and the optimised values of $\zc$, $C_{H_w}$, and $\Pi$ for these datasets are listed in appendix~\ref{apx:B}.

For some datasets, the boundary-layer thickness $\delta$ cannot be determined using the method of \citet{lozier2025defining}, due to the unavailability of skewness data.
In such cases, the composite profile fitting procedure still provides optimised values of $\zc$ and $\Pi$, together with an effective outer length scale $\delta/C_{H_w}$.
The boundary-layer thickness $\delta$ can then be estimated by assuming $C_{H_w}=1$, or alternatively $C_{H_w}$ can be estimated by assuming $\delta=1.25\delta_{99}$ \citep[following][]{lozier2025defining}.
Despite this uncertainty, the fitted composite profiles continue to match the mean velocity data well, including cases with significant PG effects (figure~\ref{fig14}).

Having established the accuracy of the composite profile across a broad range of APG TBL datasets, the fitting procedure was applied to all datasets considered in this study.
The resulting parameters were then compiled to evaluate trends associated with pressure-gradient effects.

\subsection{Overlap region}
\label{sec5.1}

\begin{figure}
    \captionsetup{width=1.00\linewidth}
    \begin{center}
    \includegraphics[width=1.00\textwidth]{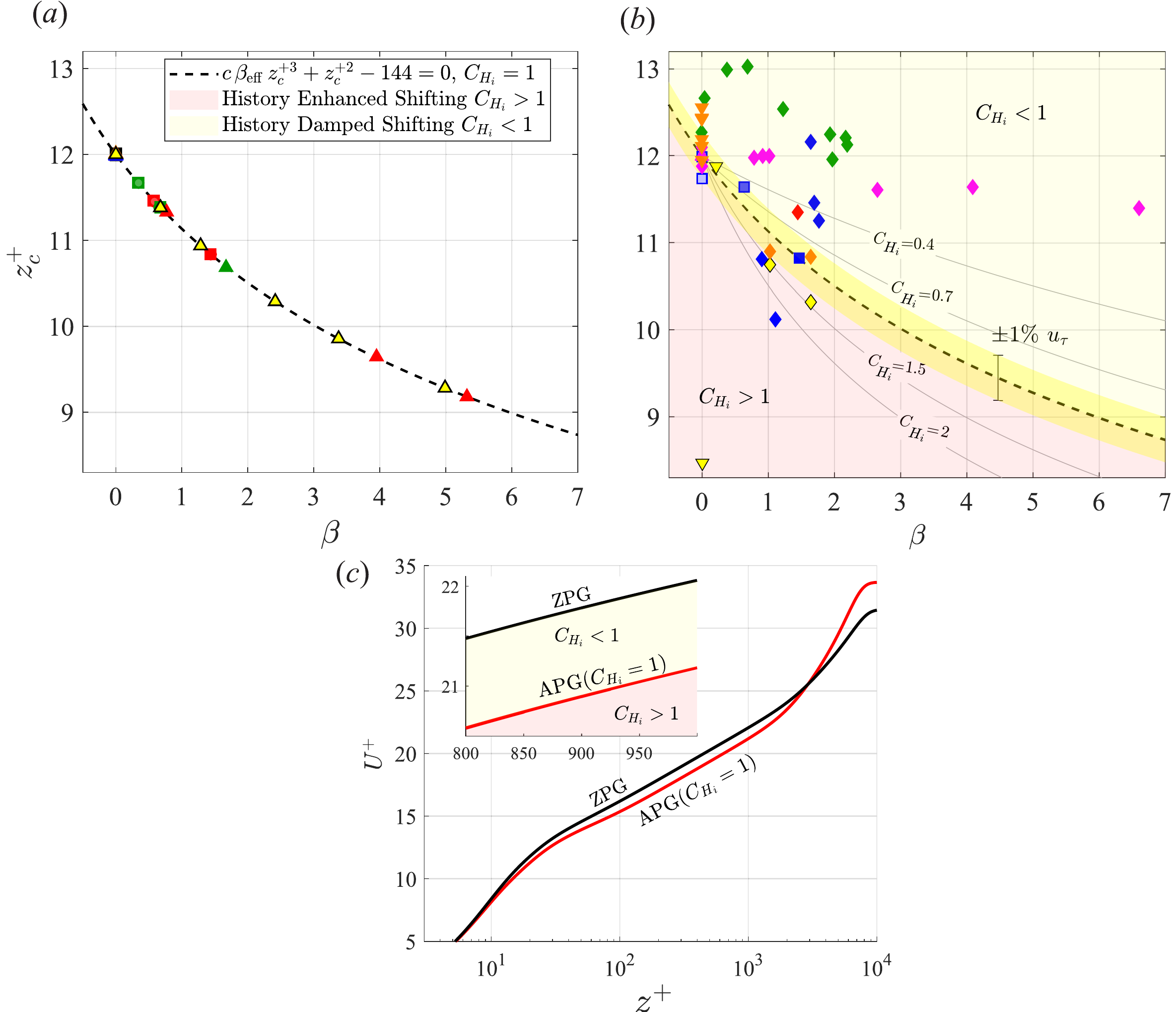}
    \end{center}
    \caption{Variation of $\zc$ with $\beta$ for (\textit{a}) cases with minimal PG history and (\textit{b}) cases with PG history effects. The dashed line represents the proposed model \eqref{eq:model}, and the grey lines represent $C_{H_i}$ levels. (\textit{c}) Schematic of the mean velocity profile for a reference high-$\re$ ZPG and an APG dataset depicting `history-enhanced' and `history-damped' shifting behaviours.}
    \label{fig9}
\end{figure}

We first examine trends in $\zc$, a key parameter governing the overlap-region behaviour of the mean velocity profile.
As discussed by \citet{nickels} and demonstrated experimentally in Part~1, $\zc$ determines the additive constant of the classical log-law and therefore controls the vertical shift of the overlap-region mean velocity profile.
This parameter is expected to depend on the pressure gradient according to the model \eqref{eq:model}.
The optimised values of $\zc$ for cases with minimal PG history are plotted as a function of $\beta$ in figure~\ref{fig9}(a).
Excellent agreement is observed between the $\zc$ values obtained from the composite-profile fitting method (where $C_{H_i}\approx 1$) and the tuned model \eqref{eq:model}, shown by the dashed black line in figures~\ref{fig9}(a,b).
Notably, $\zc$ decreases systematically with increasing APG, independent of Reynolds number, consistent with the results of Part~1.
These trends also follow the proposed $\beta$-dependent model \eqref{eq:model} more closely than the original $p_x^+$-based model \eqref{eq:pxNickles} of \citet{nickels} (not shown here).

We next consider cases with significant PG history effects, shown in figure~\ref{fig9}(b).
For these datasets, the optimised $\zc$ values deviate from the model \eqref{eq:model}, indicating the influence of PG history and resulting in $C_{H_i} \neq 1$.
Physically, $C_{H_i}$ represents the cumulative influence of upstream pressure-gradient events on the overlap-region structure of the mean velocity profile.
Values of $C_{H_i}>1$ correspond to cases where upstream PG development enhances the downward shift in the overlap region of the mean velocity  relative to that predicted by the local PG alone, whereas $C_{H_i}<1$ indicates a relative upward shift arising from the upstream PG history (figure \ref{fig9}c).
The patterns of $\zc$ observed within individual datasets reflect the unique PG histories of each case (e.g.\ differences in the $\beta(x)$ distributions arising from ramp or roof geometries in experiments).
Minor deviations from the model curve that fall within experimental uncertainty should not necessarily be interpreted as significant PG history effects.

However, some cases show clear evidence of strong PG history influence.
For example, the datasets of \citet{gungor2024} and \citet{preskett2025effects} correspond to nominally ZPG conditions ($\beta=0$ locally) but have experienced significant PG events upstream.
These cases appear as the yellow and orange downward triangles in figure~\ref{fig9}(b), lying at $\beta=0$ but with $\zc \neq 12$.
Their corresponding mean velocity profiles are shown in figure~\ref{fig14}(b), specifically case G2 from \citet{gungor2024} ($\zc=8.2$) and case Pr4 from \citep{preskett2025effects} ($\zc=12.6$), illustrating the non-canonical behaviour of these profiles despite locally ZPG conditions.

To aid interpretation of $C_{H_i}$, we consider the schematic mean velocity profiles of reference high-$\re$ ZPG and APG TBLs with minimal PG history shown in figure~\ref{fig9}(c).
The APG profile exhibits the expected downward shift relative to the ZPG profile in the overlap region.
When PG history effects are present (while maintaining the same local $\beta$), the profile may instead lie between the reference ZPG and APG cases, corresponding to $C_{H_i}<1$.
We refer to this behaviour as \textit{history-damped shifting}.
Conversely, the overlap-region shift may exceed that of the reference APG profile, corresponding to $C_{H_i}>1$, which we term \textit{history-enhanced shifting}. 
It is also noted that if the APG profile instead lies above the ZPG profile, $C_{H_{i}}<0$. 

Cases with $C_{H_i} \neq 1$ can be mapped to an effective pressure-gradient parameter $\beta_{\mathrm{eff}}$ using the measured $\zc$ and \eqref{eq:model}.
Flows with matched $\beta_{\mathrm{eff}}$ therefore exhibit similar overlap-region behaviour (i.e.\ the same additive constant $B$).
In this way, $\beta_{\mathrm{eff}}$ may be interpreted as the effective pressure gradient experienced by the boundary layer after accounting for the influence of upstream PG development.
This framework, therefore, provides a simple means of interpreting the combined influence of local and upstream PGs on the mean velocity profile.

\begin{figure}
    \captionsetup{width=1.00\linewidth}
    \begin{center}
    \includegraphics[width=1.00\textwidth]{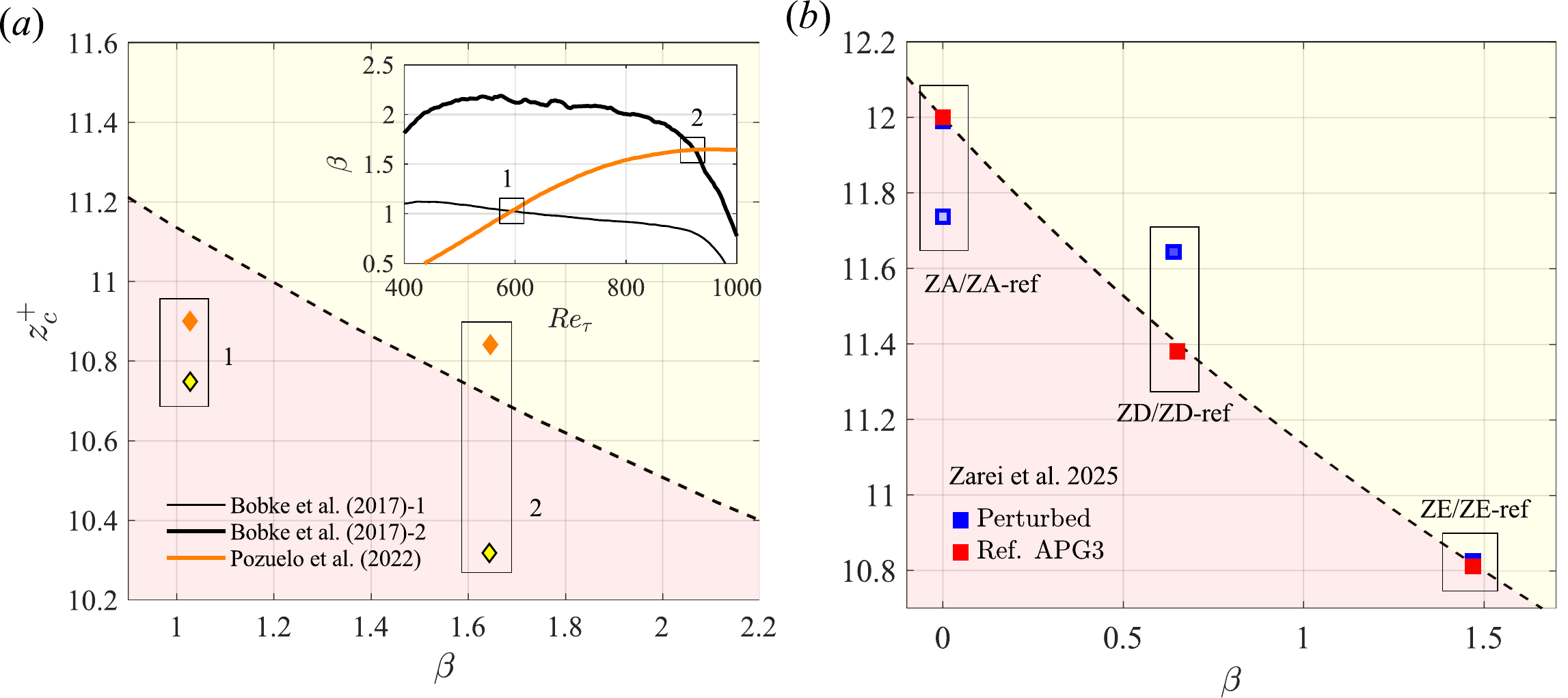}
    \end{center}
    \caption{Variation of $\zc$ with $\beta$ for cases with matched local flow conditions ($\re$ and $\beta$) but different PG histories from the datasets of (\textit{a}) \citet{pozuelo} and \citet{bobke2017history}, and (\textit{b}) \cite{zarei2026_part1}.}
    \label{ZpBetaPB}
\end{figure}

These PG history effects become clearer when considering TBLs with matched local conditions ($\re$ and $\beta$) but different PG histories.
The corresponding $\zc$ values for such cases from the numerical datasets of \citet{pozuelo} and \citet{bobke2017history}, together with the high-$\re$ experimental dataset from Part~1, are shown in figures~\ref{ZpBetaPB}(a,b).
In figure~\ref{ZpBetaPB}(a), differences in the upstream PG development (see inset $\beta(x)$ distributions) produce clear differences in $\zc$ despite matched local flow conditions.
In other words, the local and upstream PGs combine to produce different effective pressure gradients $\beta_{\mathrm{eff}}$.
For the higher-$\re$ data points (labelled 2), the case of \citet{pozuelo} exhibits history-damped shifting, whereas the case of \citet{bobke2017history} exhibits history-enhanced shifting, indicating a larger $\beta_{\mathrm{eff}}$.

In figure~\ref{ZpBetaPB}(b), the cases labelled ZA and ZD were designed to introduce measurable PG history effects relative to the reference minimal-history cases (ZA-ref.\ and ZD-ref.) from Part~1.
For ZA, an imposed upstream APG perturbation produces history-enhanced shifting relative to the reference ZPG case (ZA-ref.), corresponding to $\beta_{\mathrm{eff}}>0$.
In contrast, delayed APG development in ZD relative to the reference APG case (ZD-ref.) leads to history-damped shifting, although $\zc$ remains less than the canonical ZPG value of 12 (i.e.\ $\beta > \beta_{\mathrm{eff}} > 0$).
Finally, despite differences in upstream PG history, both ZE and ZE-ref.\ exhibit overlap-region shifts consistent with the local $\beta$, following \eqref{eq:model}.

Together, these observations demonstrate that the composite profile, fitting procedure and PG-history framework proposed here provide a consistent way to describe the combined influence of local and upstream PGs across a wide range of APG TBLs.
While the parameter $C_{H_i}$ characterises PG-history effects on the overlap-region velocity shift, the wake region can exhibit additional history-dependent behaviour.
We therefore next examine how PG history influences the wake structure through the parameter $C_{H_w}$.

\subsection{Wake region}
\label{sec5.2}

\begin{figure}
    \captionsetup{width=1.00\linewidth}
    \begin{center}
    \includegraphics[width=0.95\textwidth]{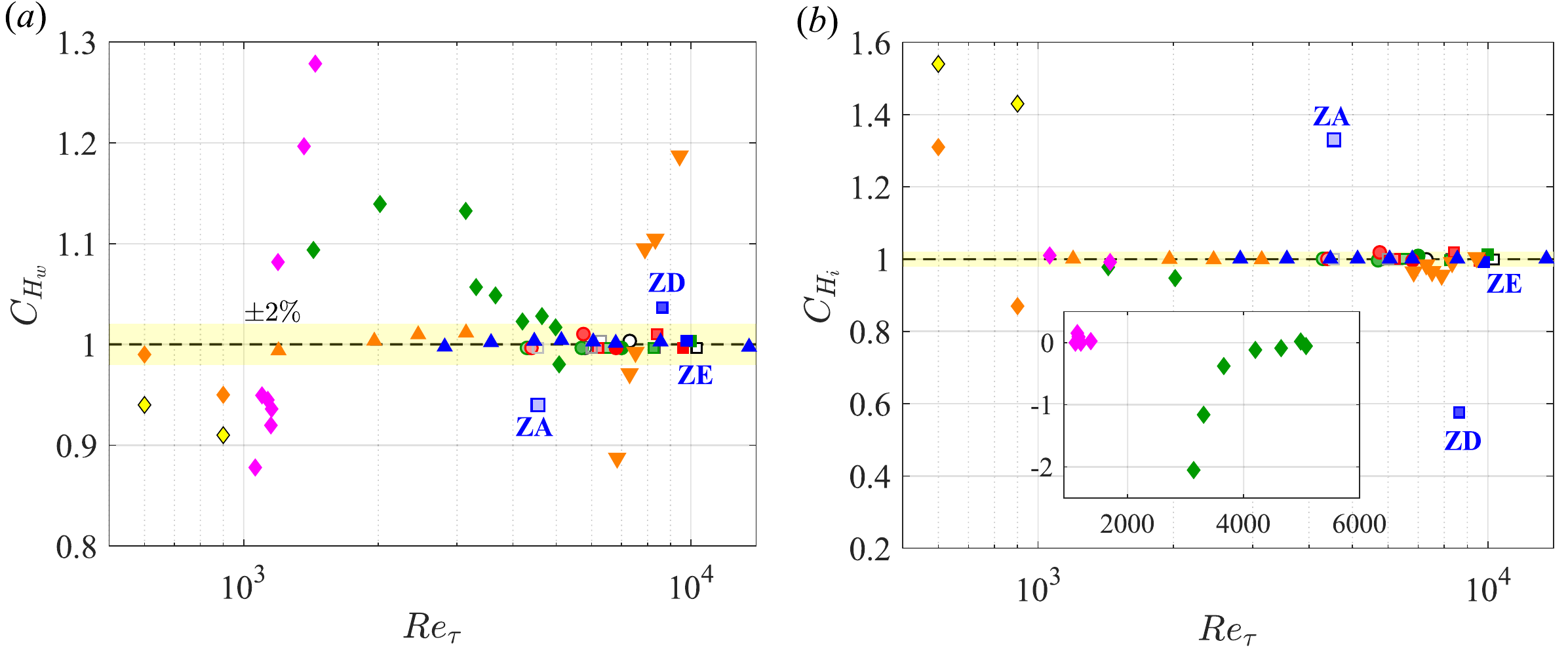}
    \end{center}
    \caption{Variation of \textit{(a)} $C_{H_w}$ and \textit{(b)} $C_{H_i}$ for all datasets considered here.}
    \label{PgCh}
\end{figure}

The wake region can also be strongly influenced by both local and upstream pressure gradients \citep{bobke2017history, gungor2024, preskett2025effects}.
In contrast to the inner and overlap regions, which adjust relatively rapidly to changes in the local flow conditions, the wake region evolves over longer streamwise development lengths and therefore retains a stronger memory of upstream PG events.
This behaviour motivates the introduction of the wake-history parameter $C_{H_w}$ in the composite profile formulation, which captures history-dependent stretching of the wake region mean velocity profile.
Consequently, it is challenging to formulate a general wake function that accounts simultaneously for a wide range of local APGs and PG-history effects while also incorporating an independent measure of $\delta$.
To demonstrate the robustness of the present composite formulation, both PG history parameters (i.e. $C_{H_w}$ and $C_{H_i}$) obtained from a subset of the datasets considered here are plotted as a function of $\re$ in figure~\ref{PgCh}.

For cases with minimal PG history, including both experimental and numerical ZPG and APG datasets, the values of $C_{H_w} \approx C_{H_i} \approx 1$ (within a 2\% margin), largely independent of $\re$ and $\beta$.
This behaviour is consistent with previous studies, such as \citet{krug2017} and \citet{chauhan2009criteria}, which employed constant coefficients in their wake functions, although these were primarily valid for `well-behaved’ TBLs.

When PG history effects are present, deviations appear in both the wake parameter ($C_{H_w}$) and the inner/overlap parameter ($C_{H_i}$).
This behaviour is evident for the cases ZA and ZD discussed in section~\ref{sec5.1}.
In some datasets, the influence of PG history is even more pronounced in $C_{H_w}$ than in $C_{H_i}$, for example, the orange downward triangles corresponding to \citet{preskett2025effects}.
This observation highlights the importance of also considering $C_{H_w}$ when characterising PG history effects.

These results, therefore, justify the inclusion of $C_{H_w}$ as a parameter in the present composite formulation.
Without this parameter, the wake function cannot simultaneously reproduce the observed variations in the wake region mean velocity profiles across datasets with different PG histories, leading to systematic deviations in the wake region of up to $\pm20\%$ in the effective wake scaling.

Further, the results also highlight the importance of incorporating an independent measure of $\delta$, since the wake function can scale with both $C_{H_w}$ and $\delta$ through $\eta_{\rm eff} = 1.455\ C_{H_w}\ z/\delta$. 
This observation is consistent with \citet{lozier2025defining}, who showed that the ratio $\delta/\delta_{99}$ (effectively a measure of wake stretching) remains nearly constant for TBLs with minimal PG history but deviates when PG history effects are present.

Finally, it should be noted that variations in tripping conditions or surface roughness perturbations can also influence the history parameters $C_{H_w}$ and $C_{H_i}$, as discussed further in appendix~\ref{apx:A}.

\section{Utilities of the new composite profile}
\label{sec:utilities}

In addition to the framework for characterising PG history effects described in section~\ref{secQuality}, and the broader applications of well-resolved mean velocity profiles in resolvent analysis or Reynolds-stress modelling, the new composite profile also provides practical utilities for analysing previously published and future APG TBL datasets.

\subsection{Revisiting previous datasets}

Accurate measurement of $\ut$ and/or $\delta$ is often challenging in APG TBL experiments, especially for those at high $Re_\tau$.
Accordingly, one of the primary utilities of the proposed composite profile is to estimate these key boundary-layer characteristics from datasets in which they are not directly measured.
To achieve this, the curve-fitting procedure described in section~\ref{secFit} is applied using the composite profile formulation together with \eqref{eq:model}.
In this case, the inputs consist of the non-normalised mean velocity profile ($U$ and $z$), the streamwise pressure gradient ($dP/dx$), and the fluid properties ($\nu$ and $\rho$).
The PG history parameters must also be specified; for well-behaved TBLs, these typically satisfy $C_{H_i}=C_{H_w}=1$.
The fitting procedure then provides optimised values of $\ut$, $\Pi$, and $\delta$, from which other boundary-layer properties may be determined.

\begin{figure}
    \captionsetup{width=1.00\linewidth}
    \begin{center}
    \includegraphics[width=1.0\textwidth]{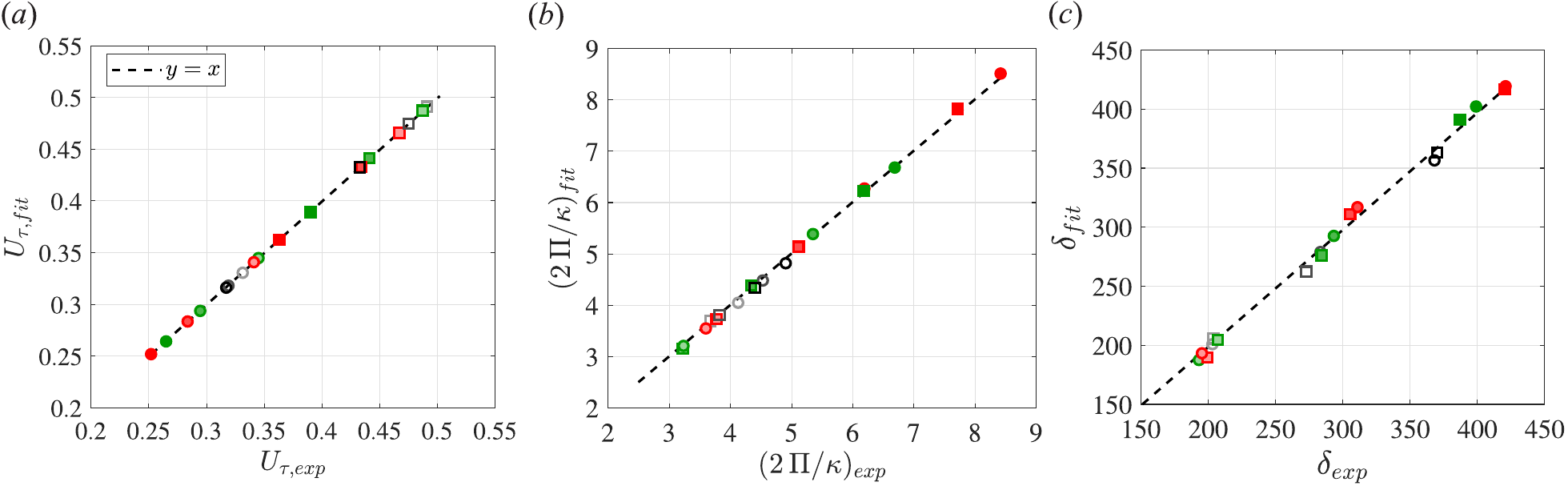}
    \end{center}
    \caption{Comparison between fitted and directly measured values of (\textit{a}) the friction velocity $\ut$, (\textit{b}) the wake strength $2\Pi/\kappa$, and (\textit{c}) the TBL thickness $\delta$ for the datasets from Part 1.}
    \label{fig10}
\end{figure}

To evaluate the accuracy of this approach, the method is applied to the datasets from Part~1, which are characterised by minimal PG history ($C_{H_i} \approx C_{H_w} \approx 1$) and for which $\ut$ and $\delta$ were independently measured.
The comparison is shown in figure~\ref{fig10}.
The values obtained from the composite-profile fitting procedure agree closely with the experimentally measured values and fall within the range of typical experimental uncertainties for all three quantities considered.

\begin{figure}
    \captionsetup{width=1.00\linewidth}
    \begin{center}
    \includegraphics[width=0.80\textwidth]{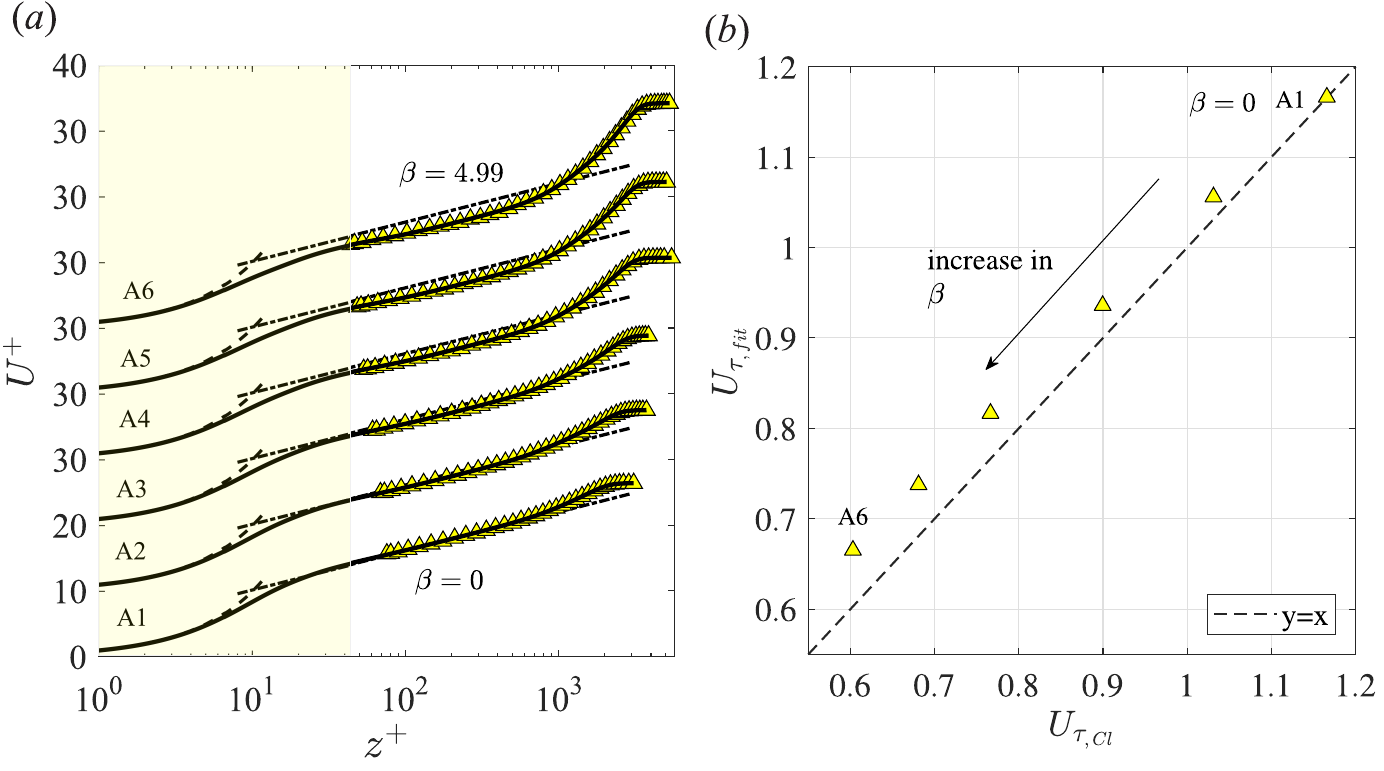}
    \end{center}
    \caption{(\textit{a}) Experimental mean velocity data and corresponding composite profiles from \citet{maruvsic1995wall}. The yellow-shaded region indicates the near-wall extent, which was not measured experimentally. (\textit{b}) Comparison of $\ut$ values obtained from the Clauser fitting method and the composite profile. Dashed lines represent \eqref{eq:closeWall} and dash-dotted lines represent \eqref{eq:loglaw}.}
    \label{fig12}
\end{figure}

This demonstrates the utility of the composite profile as a tool for revisiting datasets in which direct measurements of $\ut$ were not reported.
As an example, the method is applied to the dataset of \citet{maruvsic1995wall}, where $\ut$ was originally estimated using the Clauser method and near-wall measurements were limited.
Despite the absence of near-wall data, the composite profile matches the experimental velocity profiles well (figure~\ref{fig12}a).
Comparing the resulting friction velocities with those obtained using the Clauser method (figure~\ref{fig12}b) reveals clear differences for cases with the highest $\beta$, while the ZPG case shows good agreement.
This behaviour is expected, since the Clauser method is reliable primarily for canonical TBLs with negligible pressure gradients.

The optimised parameters obtained from this procedure are listed in table~\ref{tab:Alldatasests}.
In addition to providing improved estimates of the friction velocity, the composite profile also yields a well-resolved representation of the near-wall velocity profile, which was experimentally inaccessible for these cases (yellow-shaded region in figure~\ref{fig12}b).

\subsection{Improving accuracy of the indicator function}

Another challenge, particularly for single-point experimental hot-wire measurements, is computing wall-normal velocity derivatives and, in particular, the indicator function \eqref{eq:Indicator}.
Experimental noise, small measurement errors and sparse wall-normal resolution make it difficult to draw firm conclusions about the existence of a logarithmically scaled region or to identify its start and end points.
Similarly, numerical datasets require very high wall-normal resolution to accurately resolve the indicator function, making them computationally expensive at high $\re$ \citep{nagib2024utilizing}.

In contrast, the composite profile provides an analytical representation of the mean velocity profile that can be differentiated analytically, yielding a smooth expression for the indicator function.
For example, in the overlap region, the indicator function can be written as
\begin{equation}
    z^+ \frac{dU^+}{dz^+} = \frac{\gamma z^+}{\kappa} \cdot \frac{0.6 \, \xi^5 \, Re_\tau (1 + \eta_{\rm{eff}}^6) - \eta_{\rm{eff}}^5 \, \zc (1 + \xi^6)}{\zc \, Re_\tau \, (1 + \xi^6)(1 + \eta_{\rm{eff}}^6)};\quad\xi = \frac{0.6 z^+}{\zc},\quad \gamma = \sqrt{1 + p_x^+ \zc}
    \label{eq:Indicator}
\end{equation}
In the high-$\re$ limit, this expression reduces to $z^+\frac{dU^+}{dz^+} = 1/\kappa$, consistent with the classical log-law behaviour expected in the overlap region of high-$\re$ TBLs.

\begin{figure}
    \captionsetup{width=1.00\linewidth}
    \begin{center}
    \includegraphics[width=0.90\textwidth]{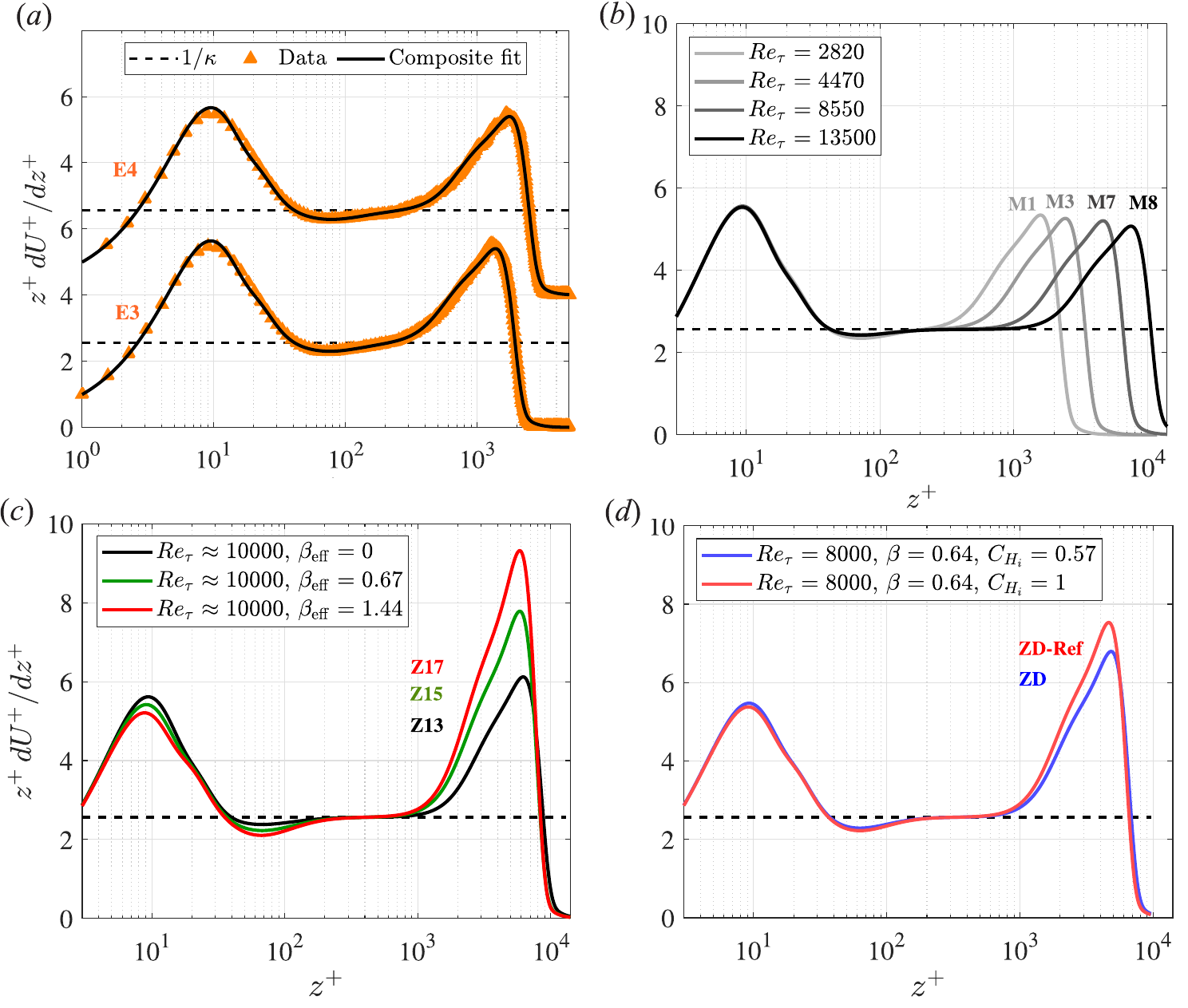}
    \end{center}
    \caption{Indicator functions computed from composite profiles fitted to (\textit{a}) the low-$\re$ ZPG LES dataset of \citet{eitel2014simulation}, (\textit{b}) the high-$\re$ ZPG experimental dataset of \citet{marusic2015}, and the high-$\re$ moderate-APG experimental dataset from Part 1 with (\textit{c}) minimal PG histories or (\textit{d}) imposed PG history effects. Horizontal dashed lines represent $1/\kappa$.}
    \label{fig15}
\end{figure}

The indicator function also provides a useful criterion for assessing the quality of the composite-profile fit.
Figure~\ref{fig15}(a) compares the indicator function computed directly from the well-resolved ZPG LES of \citet{eitel2014simulation} with that obtained from the corresponding composite profile, demonstrating excellent agreement.

Further insight can be obtained by considering indicator functions derived from fitted high-$\re$ experimental datasets.
For example, the plateau associated with the overlap region becomes increasingly pronounced as $\re$ increases, reflecting the growing separation between the inner and wake regions and therefore a larger overlap-region extent.
This behaviour is shown in figure~\ref{fig15}(b) for the high-$\re$ ZPG experimental dataset of \citet{marusic2015}.
For the highest-$\re$ case (M8), the pronounced plateau corresponding to $1/\kappa$ indicates a clear overlap region following classical logarithmic scaling.
In contrast, at lower $\re$ (e.g. case M1 in figure~\ref{fig15}b and the cases in figure~\ref{fig15}a), a well-defined plateau is not observed.

The effect of increasing APG on the indicator function at approximately $\re=10,000$ is illustrated in figure~\ref{fig15}(c) using experimental datasets from Part~1 with $\beta_{\rm eff}=$ 0, 0.66 and 1.44.
For moderate APGs at high $\re$, the indicator function still exhibits a clear plateau, indicating that the overlap region follows logarithmic scaling. 
However, as $\beta_{\rm eff}$ increases, the wake strengthens and the plateau terminates earlier, reducing the logarithmic extent of the overlap region. 
Increasing $\beta_{\rm eff}$ also reduces the near-wall velocity gradient, decreasing the first peak of the indicator function, while simultaneously increasing the buffer-region overshoot, which deepens the local minimum (figure~\ref{fig15}c).

Figure~\ref{fig15}(d) further illustrates the influence of PG history on the indicator function and the extent of the overlap region.
The case corresponding to $C_{H_i}=0.57$ exhibits a slightly longer overlap region than the case with $C_{H_i}=1$, despite having matched $\re$ and $\beta$.

\begin{figure}
    \captionsetup{width=1.00\linewidth}
    \begin{center}
    \includegraphics[width=0.9\textwidth]{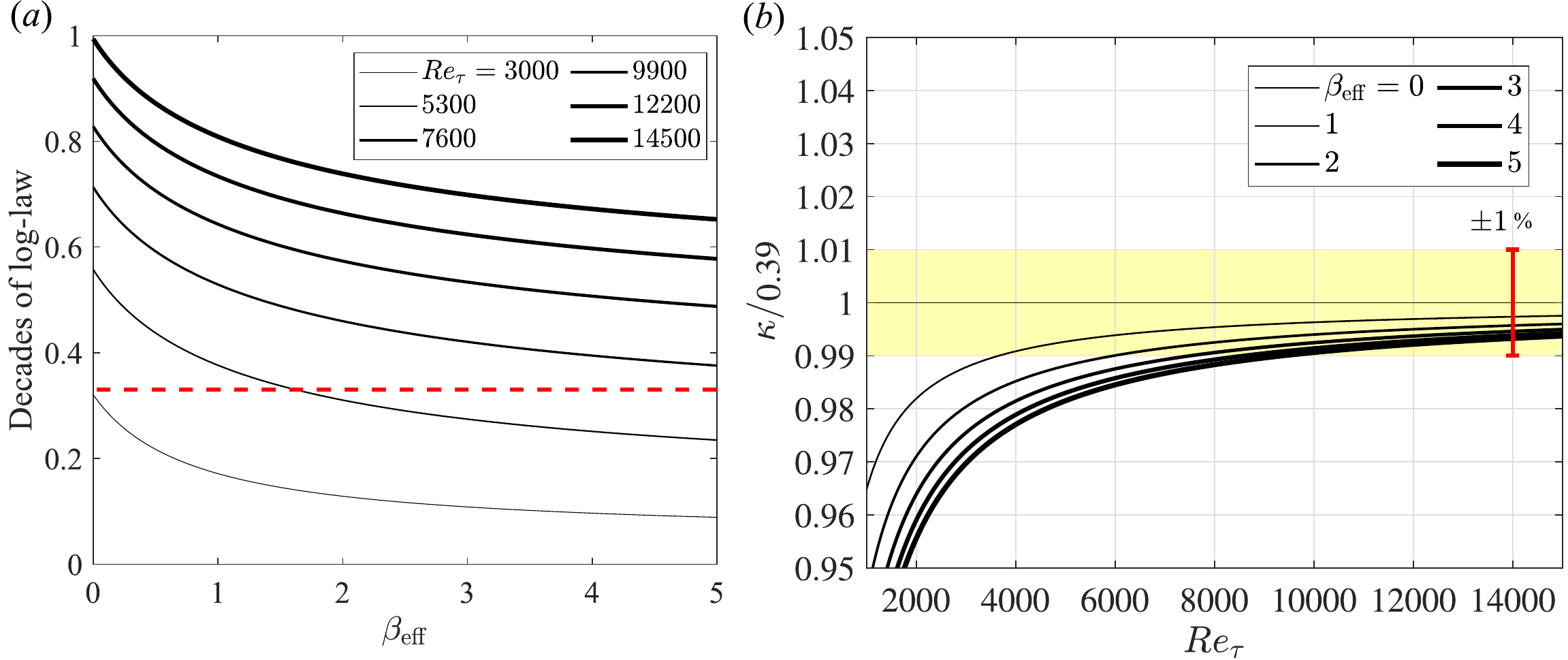}
    \end{center}
    \caption{Variation in (a) the decades of $\zp$ which follow the classical log-law and (b) the value of $\kappa$ as a function of $\re$ and $\beta_{\rm eff}$. The red dashed line indicates 0.33 decades of log-law.}
    \label{DecadesLogLaw}
\end{figure}

The composite profile and the associated indicator function can also be used to estimate the number of decades of $\zp$ over which logarithmic scaling occurs in the overlap region by measuring the plateau extent within a given tolerance (e.g. $1/\kappa \pm 1\%$).
Predicted variations in the number of decades of log-law are shown as a function of $\beta_{\rm eff}$ and $\re$ in figure~\ref{DecadesLogLaw}(a).
In this analysis the approximations of \citet{perry_streamwise_2002} were employed: $\Pi = 0.68(\beta + 0.5)^{0.75}$ and $\delta / \delta^+ = 7.912 / (1 + 0.1861 \beta)$, which are valid for $\beta < 6$.
The results demonstrate competing but expected trends: the logarithmic extent decreases with increasing $\beta$ at fixed $\re$, but increases with $\re$ at fixed $\beta$.
The red dashed line denotes a minimum 1/3 of a decade in $z^+$, roughly corresponding to the predicted logarithmic extent of a ZPG TBL at $\re=3000$.
Such predictions provide useful guidelines for selecting experimental parameter ranges for future investigations of logarithmic scaling in APG TBLs.

The composite profile also predicts the slope of the mean velocity profile in the overlap region, i.e.\ the von Kármán coefficient.
According to \eqref{eq:Presentoverlap}, $\kappa$ varies as $\kappa/0.39 = 1/\sqrt{1 + p_x^{+} \zc}$.
Predicted variations of $\kappa$ with $\beta_{\rm eff}$ are shown as a function of $\re$ in figure~\ref{DecadesLogLaw}(b).
At low $\re$, $\kappa$ varies significantly with $\beta_{\rm eff}$, whereas for $\re \gtrsim 10,000$ the curves approach $\kappa=0.39$ (within typical experimental uncertainty) regardless of $\beta_{\rm eff}$.
This indicates that the $p_x^+$ terms in \eqref{eq:PresentInner} and \eqref{eq:Presentoverlap} become negligible at sufficiently high $\re$, consistent with \eqref{eg:overlapHigh}.
It also suggests that the hierarchy of attached eddies in the overlap region remains largely unchanged at high $\re$, even under moderate APGs or PG history effects, consistent with \citet{woodcock2015statistical}.

\begin{figure}
    \captionsetup{width=1.00\linewidth}
    \begin{center}
    \includegraphics[width=0.5\textwidth]{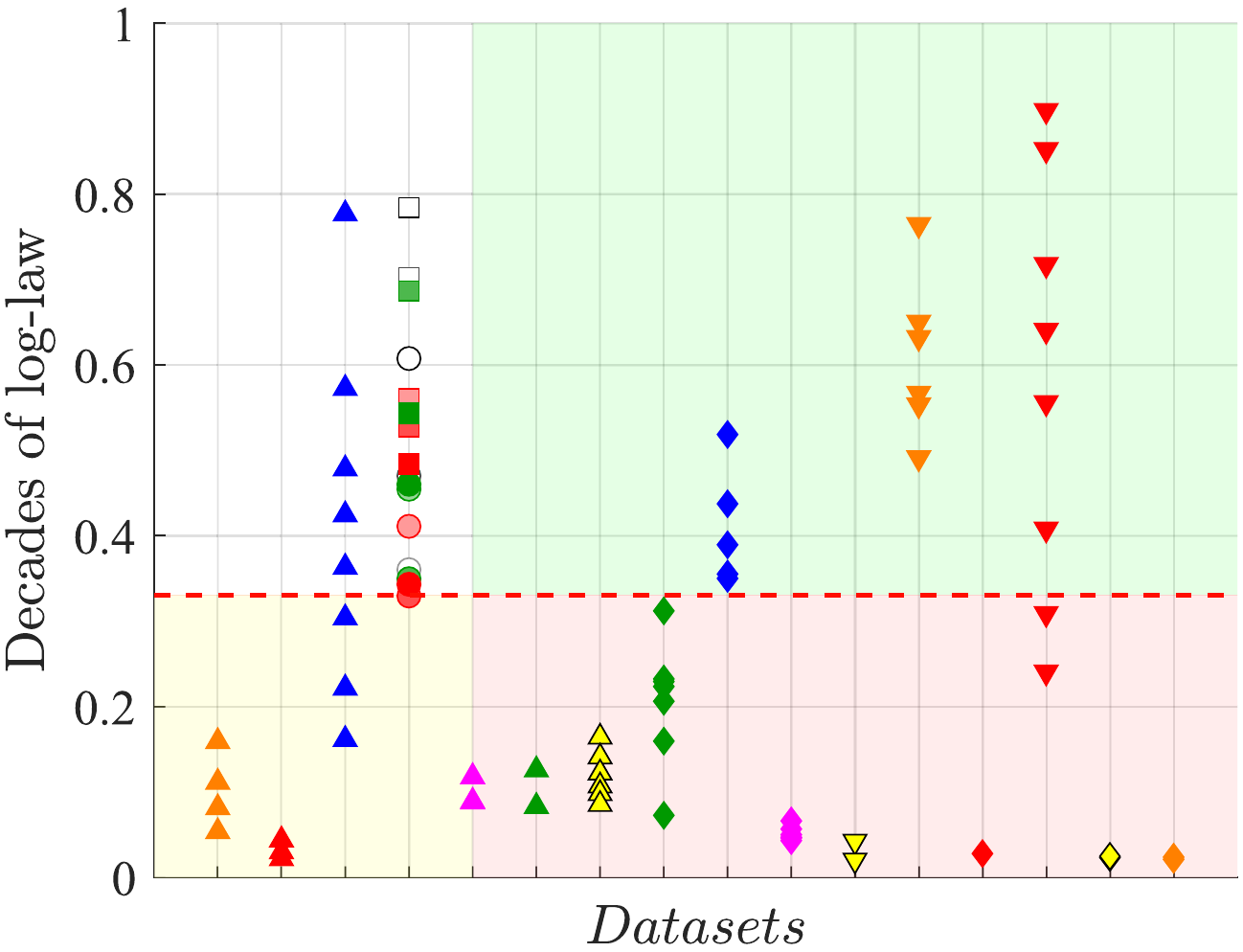}
    \end{center}
    \caption{Decades of log-law for the datasets listed in table~\ref{tab:database}. The red dashed line indicates the threshold of 0.33 decades of log-law. Cases below this threshold are highlighted in yellow, datasets with noted PG history effects in the overlap region are highlighted in green and cases with both conditions are highlighted in red.}
    \label{DecadesLogLawAll}
\end{figure}

Figure~\ref{DecadesLogLawAll} shows the estimated wall-normal extent of the log-law, in decades of $z^+$, for the datasets in table~\ref{tab:database}, based on indicator functions derived from the fitted composite profiles.
The red dashed line again denotes the approximate threshold of 1/3 of a decade in $z^+$ required to clearly identify logarithmic scaling.
Cases below this threshold are highlighted in yellow, datasets with previously noted PG history effects in the overlap region are highlighted in green, and cases satisfying both conditions are highlighted in red.
This comparison highlights the uniqueness of the dataset presented in Part~1, which simultaneously maintains a significant overlap region (high $\re$ and moderate $\beta$) and controlled PG history effects, enabling a clear investigation of logarithmic scaling in APG TBLs.

A further utility of the composite profile is that it provides an analytical expression from which inflection points ($d^{2}U/dz^{2}=0$) can be identified.
This is useful for testing the framework of \citet{schatzman2017experimental}, which links APG-induced energisation of wake-region motions to a shear instability associated with an inflectional mean velocity profile.
The availability of an analytical composite profile, therefore, facilitates identification of these inflection points and further assessment of this mechanism for APG TBL dynamics.

\section{Conclusions}

This study presents a novel and practical composite profile formulation, \eqref{eq:PresentInner}–\eqref{eq:Presentwake}, for the mean streamwise velocity in adverse-pressure-gradient (APG) turbulent boundary layers (TBLs), with three physically meaningful free parameters: $\zc$, $C_{H_w}$ and $\Pi$.
The formulation is motivated by the results of Part~1 and strategically modifies the earlier composite profile of \citet{nickels}, originally developed for generic low-$\re$ pressure-gradient TBLs, with the aim of extending its applicability to a broader range of Reynolds numbers and PG conditions.

Several key modifications were introduced.
First, the wake function was reformulated to be consistent with an independent and physically motivated definition of the boundary-layer thickness \citep{lozier2025defining}, together with the introduction of a wake-history parameter $C_{H_w}$ that accounts for pressure-gradient history effects in the wake region.
Second, a velocity-overshoot function was incorporated in the inner region, consistent with the approaches of \citet{monkewitz2007self} and \citet{chauhan2009criteria}.
Third, the relationship between streamwise PGs and the sublayer thickness $\zc$ originally proposed by \citet{nickels} was reformulated through \eqref{eq:model} in light of recent high-$\re$ APG TBL studies.

The proposed composite profile was evaluated using a compilation of APG TBL datasets spanning more than three decades of experimental and numerical investigations, with particular emphasis on the high-$\re$ dataset presented in Part~1.
A nonlinear least-squares curve-fitting procedure, implemented in the accompanying \href{https://cocalc.com/share/public_paths/cd3b9c28f1e14b1de432482bc8009c461a69cfad}{notebooks}, was used to optimise the free parameters and fit the mean velocity profiles across all cases considered.
The resulting fits demonstrate that the composite profile provides an accurate representation of the mean velocity structure across a wide range of Reynolds numbers, pressure-gradient magnitudes and PG-history conditions.

The optimised parameters further enable the development of a framework for interpreting PG-history effects.
In particular, the wake-history parameter $C_{H_w}$ quantifies the influence of upstream PG development on the wake-region structure, while the effective pressure-gradient parameter $\beta_{\rm eff}$ (or equivalently $C_{H_i}$), derived from $\zc$, characterises the combined influence of local and upstream PGs on the overlap-region velocity shift.
These parameters provide a practical criterion for identifying `well-behaved' APG TBLs, for which $C_{H_w} \approx 1$ and $\beta=\beta_{\rm eff}$, and allow cases with PG history to be classified in terms of `history-enhanced' or `history-damped' modifications to the mean velocity profile.

The composite profile also provides several practical utilities for analysing APG TBL datasets.
In particular, it enables estimation of the friction velocity $\ut$ and boundary-layer thickness $\delta$ in cases where these quantities cannot be measured accurately, provided the PG history is minimal or well characterised.
For the dataset from Part~1, the values of $\ut$ and $\delta$ obtained from this approach agree with direct experimental measurements within typical experimental uncertainty.
The analytical form of the composite profile additionally allows smooth differentiation of the mean velocity profile, facilitating evaluation of the indicator function and identification of inflection points, which are often difficult to obtain from sparse or noisy experimental data.
The formulation also enables predictions regarding the dependence of the logarithmic-region extent and the von Kármán coefficient on Reynolds number and pressure-gradient strength, providing useful guidance for future experimental investigations.

Beyond providing an accurate analytical representation of the mean velocity profile, the formulation also yields several physical insights.
The results demonstrate that pressure-gradient history modifies the mean velocity profile in distinct ways across the boundary layer, influencing both the overlap-region velocity shift and the wake-region structure through separate mechanisms characterised by $C_{H_i}$ and $C_{H_w}$.
The concept of an effective pressure gradient $\beta_{\rm eff}$ provides a simple framework for interpreting the combined influence of local and upstream PGs on overlap scaling.

In summary, the composite profile formulation introduced here provides a unified and physically interpretable description of APG TBL mean velocity profiles across a wide range of Reynold numbers and pressure-gradient conditions.
While the formulation is broadly applicable to pressure-gradient TBLs, it has not yet been tested extensively for favourable pressure gradients or flows approaching separation.
Additional high-fidelity experimental and numerical datasets, particularly at high Reynolds numbers, will therefore be valuable for further refining the model.
Extending the framework to rough-wall TBLs would likely require additional parameters, and further investigations of rough-wall pressure-gradient boundary layers are needed to support such developments.

\backsection[Acknowledgements]{\noindent The authors gratefully acknowledge funding from the Office of Naval Research (ONR) and ONR Global; Grant No. N62909-23-1-2068. R. Deshpande also acknowledges financial support from the Melbourne Postdoctoral Fellowship awarded by the University of Melbourne.}

\backsection[Declaration of Interests]{The authors report no conflict of interest.}

\appendix
\section{Fitting results for all datasets considered in the present study}
\label{apx:B}

\renewcommand{\arraystretch}{1.0}
\setlength{\tabcolsep}{3pt}     
\scriptsize

\begin{longtable}{c c c >{\columncolor{yellow!20}}c >{\columncolor{yellow!20}}c >{\columncolor{Green!10}}c >{\columncolor{Green!10}}c >{\columncolor{Green!10}}c c c c}
    \hline
    \cellcolor{white}Dataset & \cellcolor{white}Case no & \cellcolor{white}$\beta$ & \cellcolor{white}$\re$ & \cellcolor{white}$p_x^+ \times10^{4}$ & \cellcolor{white}$\Pi$ & \cellcolor{white}$\zc$ & \cellcolor{white}$C_{H_w}$ & \cellcolor{white}$C_{H_i}$ & \cellcolor{white}$\beta_{\rm eff}$ & \cellcolor{white}$\delta$ \\
    \hline
    \multicolumn{11}{c}{\textit{Datasets with all relevant TBL properties known.}} \\ 
    \hline
    \endfirsthead

    \multicolumn{11}{c}{\textit{(Continued from previous page)}} \\
    \hline
    \cellcolor{white}Dataset & \cellcolor{white}Case no & \cellcolor{white}$\beta$ & \cellcolor{white}$\re$ & \cellcolor{white}$p_x^+ \times10^{4}$ & \cellcolor{white}$\Pi$ & \cellcolor{white}$\zc$ & \cellcolor{white}$C_{H_w}$ & \cellcolor{white}$C_{H_i}$ & \cellcolor{white}$\beta_{\rm eff}$ & \cellcolor{white}$\delta$ \\
    \hline
    \endhead

    \hline \multicolumn{11}{r}{\textit{Continued on next page}} \\
    \endfoot

    \hline
    \endlastfoot

\citet{zarei2026_part1} & ZA	& 0    & 4550	& 0	    & 0.6	&11.74	&0.94	&1.33	&0.27	&132.8	\\
& ZA-Ref.& 0    & 4090	& 0	    & 0.45	&12  	&1	    &1	    &0	    &124	\\
& ZB	& 0	   & 5840	& 0     & 0.66	&12	    &1	    &1	    &0	    &200	\\
& ZC	& 0	   & 7440	& 0 	& 0.79	&12  	&1	    &1	    &0	    &268	\\
& ZD	& 0.64 & 8000	& 6.5	& 1	    &11.64	&0.96	&0.58	&0.37	&296	\\
& ZD-Ref.& 0.65 & 7850	& 6.6	& 1.15	&11.36	&1	    &1	    &0.65	&295	\\
& ZE	& 1.47 & 9790	& 9.8	& 1.51	&10.83	&1	    &1	    &1.46	&434	\\
& ZE-Ref.& 1.46 & 10000	& 10	& 1.5	&10.8	&1	    &1	    &1.46	&440	\\

\\
& Z1 & 0 & 6380 & 0 & 0.71 & 12.00 & 1 & 1 & 0 & 204 \\
& Z2 & 0 & 4340 & 0 & 0.79 & 12.00 & 1 & 1 & 0 & 203 \\
& Z3 & 0 & 6610 & 0 & 0.6 & 12.00 & 1 & 1 & 0 & 207 \\
& Z4 & 0 & 4310 & 0 & 0.61 & 12.00 & 1 & 1 & 0 & 193 \\
& Z5 & 0 & 6150 & 0 & 0.71 & 12.00 & 1 & 1 & 0 & 200 \\
& Z6 & 0 & 4390 & 0 & 0.68 & 12.00 & 1 & 1 & 0 & 195 \\
& Z7 & 0 & 8300 & 0 & 0.73 & 12.00 & 1 & 1 & 0 & 273 \\
& Z8 & 0 & 5800 & 0 & 0.86 & 12.00 & 1 & 1 & 0 & 284 \\
& Z9 & 0.34 & 8220 & 3.6 & 0.84 & 11.67 & 1 & 1 & 0.34 & 284 \\
& Z10 & 0.34 & 5650 & 4.7 & 1.03 & 11.67 & 1 & 1 & 0.34 & 293 \\
& Z11 & 0.57 & 8430 & 5.8 & 0.98 & 11.46 & 1 & 1 & 0.57 & 305 \\
& Z12 & 0.58 & 5760 & 7.5 & 1.20 & 11.45 & 1 & 1 & 0.58 & 311 \\
& Z13 & 0 & 10680 & 0 & 0.85 & 12.00 & 1 & 1 & 0 & 370 \\
& Z14 & 0 & 7340 & 0 & 0.94 & 12.00 & 1 & 1 & 0 & 368 \\
& Z15 & 0.67 & 9970 & 5.1 & 1.19 & 11.38 & 1 & 1 & 0.67 & 387 \\
& Z16 & 0.66 & 7000 & 6.7 & 1.28 & 11.39 & 1 & 1 & 0.66 & 399 \\
& Z17 & 1.44 & 9680 & 9.8 & 1.50 & 10.84 & 1 & 1 & 1.44 & 419 \\
& Z18 & 1.44 & 6760 & 13.2 & 1.63 & 10.84 & 1 & 1 & 1.44 & 421 \\
\\
\citet{li2021experimental} & L1 & 0 & 8610 & 0 & 1.47 & 11.09 & 1.14 & 1.08 & 1.06 & 154 \\
& L2 & 0 & 8880 & 0 & 1.24 & 11.16 & 1.04 & 1.08 & 0.96 & 156 \\
& L3 & 0 & 8920 & 0 & 1.00 & 11.56 & 0.94 & 1.04 & 0.47 & 159 \\
& L4 & 0 & 10880 & 0 & 0.71 & 11.60 & 0.88 & 1.03 & 0.42 & 182 \\
& L5 & 0 & 12430 & 0 & 0.65 & 11.69 & 0.88 & 1.03 & 0.32 & 216 \\
& L6 & 0 & 14650 & 0 & 0.61 & 12.00 & 0.94 & 1.00 & 0.00 & 244 \\
& L7 & 0 & 20770 & 0 & 0.57 & 12.00 & 1.01 & 1.00 & 0.00 & 345 \\
& L8 & 0 & 23930 & 0 & 0.56 & 11.95 & 1.01 & 1.00 & 0.00 & 418 \\
\\
\citet{eitel2014simulation} & E1 & 0 & 1200 & 0 & 0.63 & 12.00 & 1.00 & 1.00 & 0.00 & 66 \\
& E2 & 0 & 1960 & 0 & 0.70 & 12.00 & 1.00 & 1.00 & 0.00 & 114 \\
& E3 & 0 & 2460 & 0 & 0.71 & 12.00 & 1.00 & 1.00 & 0.00 & 147 \\
& E4 & 0 & 3140 & 0 & 0.71 & 12.00 & 1.00 & 1.00 & 0.00 & 191 \\
\\
\citet{orlu2013} & O1 & 0 & 1970 & 0 & 0.59 & 12.00 & 1.06 & 1.00 & 0.00 & 38 \\
& O2 & 0 & 1160 & 0 & 0.61 & 12.00 & 1.00 & 1.00 & 0.00 & 34\\
& O3 & 0 & 1150 & 0 & 0.63 & 12.00 & 1.00 & 1.00 & 0.00 & 34\\
\\
\citet{marusic2015} & M1 & 0 & 2820 & 0 & 0.69 & 12.00 & 1.00 & 1.00 & 0.00 & 58 \\
& M2 & 0 & 3570 & 0 & 0.71 & 12.00 & 1.00 & 1.00 & 0.00 & 75 \\
& M3 & 0 & 4470 & 0 & 0.66 & 12.00 & 1.00 & 1.00 & 0.00 & 95 \\
& M4 & 0 & 5130 & 0 & 0.63 & 12.00 & 1.00 & 1.00 & 0.00 & 109 \\
& M5 & 0 & 6050 & 0 & 0.66 & 12.00 & 1.00 & 1.00 & 0.00 & 133 \\
& M6 & 0 & 6790 & 0 & 0.63 & 12.00 & 1.00 & 1.00 & 0.00 & 150 \\
& M7 & 0 & 8550 & 0 & 0.66 & 12.00 & 1.00 & 1.00 & 0.00 & 190 \\
& M8 & 0 & 13500 & 0 & 0.63 & 12.00 & 1.00 & 1.00 & 0.00 & 319 \\

(different tripping) & MA & 0 & 6040 & 0 & 0.06 & 12.00 & 0.85 & 1.00 & 0.00 & 119 \\
& MB & 0 & 8840 & 0 & 0.32 & 12.00 & 1.13 & 1.00 & 0.00 & 187 \\
& MC & 0 & 10280 & 0 & 0.44 & 12.00 & 1.14 & 1.00 & 0.00 & 225 \\
& MD & 0 & 11940 & 0 & 0.55 & 12.00 & 1.00 & 1.00 & 0.00 & 271 \\
\\
\citet{monty2011} & T1 & 0 & 2970 & 0 & 0.65 & 12.00 & 1.08 & 1.00 & 0.00 & 62 \\
& T2 & 1.52 & 3170 & 34.1 & 1.26 & 10.79 & 1.14 & 1.00 & 1.52 & 90 \\
\\
\citet{volino2020non} & VA & 0.00 & 1060 & 0.0 & 0.74 & 11.88 & 0.88 & 1.01 & 0.12 & 14 \\
& VB & 0.00 & 1450 & 0.0 & 0.74 & 12.10 & 1.28 & 0.99 & -0.09 & 20 \\
& V1 & 2.65 & 1130 & 118.5 & 1.31 & 11.61 & 0.94 & 0.15 & 0.41 & 18 \\
& V2 & 4.09 & 1150 & 158.2 & 1.82 & 11.64 & 0.94 & 0.09 & 0.37 & 21 \\
& V3 & 6.60 & 1150 & 222.8 & 2.53 & 11.40 & 0.92 & 0.10 & 0.65 & 25 \\
& V4 & 0.79 & 1360 & 36.6 & 1.37 & 11.98 & 1.20 & 0.02 & 0.02 & 22 \\
& V5 & 0.92 & 1190 & 38.6 & 1.94 & 12.00 & 1.08 & 0.00 & 0.00 & 24 \\
& V6 & 1.02 & 1100 & 36.4 & 2.66 & 12.00 & 0.95 & 0.00 & 0.00 & 25 \\
\\
\citet{vila2020} & SA & 0.00 & 1430 & -0.2 & 0.68 & 12.27 & 1.09 & 0.98 & -0.25 & 18 \\
& SB & 0.04 & 2020 & 1.7 & 0.86 & 12.66 & 1.14 & 0.95 & -0.56 & 27 \\
& S1 & 0.38 & 3140 & 9.5 & 1.06 & 12.99 & 1.13 & -2.05 & -0.78 & 48 \\
& S2 & 0.69 & 3310 & 14.9 & 1.17 & 13.03 & 1.06 & -1.16 & -0.80 & 54 \\
& S3 & 1.23 & 3660 & 21.8 & 1.37 & 12.54 & 1.05 & -0.38 & -0.46 & 65 \\
& S4 & 1.94 & 4200 & 27.5 & 1.59 & 12.25 & 1.02 & -0.12 & -0.23 & 81 \\
& S5 & 1.97 & 4980 & 20.9 & 2.04 & 11.96 & 1.02 & 0.02 & 0.04 & 115 \\
& S6 & 2.17 & 4650 & 25.7 & 1.93 & 12.21 & 1.03 & -0.09 & -0.19 & 101 \\
& S7 & 2.19 & 5070 & 22.3 & 2.01 & 12.13 & 0.98 & -0.06 & -0.12 & 117 \\
\\
\citet{bobke2017history} & B1 & 1.03 & 580 & 83.9 & 1.39 & 10.75 & 0.94 & 1.54 & 1.58 & 45.9 \\
& B2 & 1.64 & 900 & 80.9 & 1.96 & 10.32 & 0.91 & 1.43 & 2.36 & 97.8 \\
\\
\citet{pozuelo} & P1 &1.03 & 600 & 91.2 & 1.14 & 10.9 & 0.99 &	1.31 & 1.34 & 41.9\\
& P2 & 1.65 & 920 & 84.7 & 1.73 & 10.84 & 0.95 & 0.87 & 1.43 & 88.9\\
\\

\hline
    \multicolumn{11}{c}{\textit{Datasets where $\delta$ is not known.}} \\ 
\hline

\citet{nagano} & N1 & 0.77 &  \cellcolor{white}420 & 91.2 & 0.83 & 11.33 & 0.96 & 1.00 & 0.77 & \cellcolor{Green!10}16 \\
& N2 & 3.95 & \cellcolor{white}570 & 256.0 & 2.14 & 9.64  & 0.93 & 1.00 & 3.95 & \cellcolor{Green!10}34 \\
& N3 & 5.32 & \cellcolor{white}600 & 287.0 & 2.99 & 9.18  & 0.92 & 1.00 & 5.32 & \cellcolor{Green!10}46 \\
\\
\citet{preskett2025effects} & K1 & 0.00 & \cellcolor{white}6840 & 0.0 & 0.29 & 12.44 & 1.13 & 0.96 & -0.38 & \cellcolor{Green!10}93 \\
& K2 & 0.00 &  \cellcolor{white}7290 & 0.0 & 0.48 & 12.19 & 1.03 & 0.98 & -0.18 & \cellcolor{Green!10}104 \\
& K3 & 0.00 &  \cellcolor{white}7520 & 0.0 & 0.69 & 12.43 & 1.01 & 0.97 & -0.38 & \cellcolor{Green!10}113 \\
& K4 & 0.00 &  \cellcolor{white}7890 & 0.0 & 1.04 & 12.56 & 0.91 & 0.96 & -0.48 & \cellcolor{Green!10}125 \\
& K5 & 0.00 &  \cellcolor{white}8330 & 0.0 & 1.21 & 12.11 & 0.91 & 0.99 & -0.11 & \cellcolor{Green!10}135 \\
& K6 & 0.00 &  \cellcolor{white}9440 & 0.0 & 1.80 & 11.95 & 0.84 & 1.00 & 0.05 & \cellcolor{Green!10}159 \\
\\
\citet{romero2022} & R1 & 0.90 &  \cellcolor{white}7100 & 10.7 & 0.64 & 10.81 & 0.78 & 1.64 & 1.48 & \cellcolor{Green!10}417 \\
& R2 & 1.11 &  \cellcolor{white}7120 & 12.3 & 0.69 & 10.12 & 0.67 & 2.50 & 2.77 & \cellcolor{Green!10}434 \\
& R3 & 1.65 &  \cellcolor{white}7130 & 18.5 & 0.93 & 12.16 & 0.81 & -0.09 & -0.15 & \cellcolor{Green!10}464 \\
& R4 & 1.70 &  \cellcolor{white}7640 & 18.3 & 1.01 & 11.46 & 0.67 & 0.34 & 0.58 & \cellcolor{Green!10}528 \\
& R5 & 1.77 &  \cellcolor{white}7770 & 17.1 & 1.11 & 11.26 & 0.74 & 0.47 & 0.84 & \cellcolor{Green!10}567 \\
\\
\citet{yoon} & Y1 & 1.45 &  \cellcolor{white}834 & 10.0 & 1.63 & 11.35 & 0.99 & 0.49 & 0.71 & \cellcolor{Green!10}44 \\
\\
\citet{gungor2024} & G1	& 0.22 &  \cellcolor{white}860 & 15.9 & 0.63 & 11.88 & 1.01 &	0.55 & 0.12 & \cellcolor{Green!10}6\\
& G2 & 0.01 &  \cellcolor{white}1520 & 28.4 & 5.54 & 8.47 & 0.98 & 1.42 & 8.21 & \cellcolor{Green!10}33\\

\hline
    \multicolumn{11}{c}{\textit{Dataset where $\delta$ and $\ut$ are not known.}} \\ 
\hline

\citet{maruvsic1995wall} & A1 & 0.00 & \cellcolor{white}2840 & \cellcolor{white}0.0 & 0.53 & 12.00 & \cellcolor{white}1.00 & 1.00 & 0.00 & \cellcolor{Green!10}39 \\
& A2 & 0.68 & \cellcolor{white}3210 & \cellcolor{white}10.0 & 0.76 & 11.38 & \cellcolor{white}1.00 & 1.00 & 0.68 & \cellcolor{Green!10}48 \\
& A3 & 1.28 & \cellcolor{white}3590 & \cellcolor{white}20.0 & 1.03 & 10.94 & \cellcolor{white}1.00 & 1.00 & 1.28 & \cellcolor{Green!10}61 \\
& A4 & 2.42 & \cellcolor{white}3930 & \cellcolor{white}30.0 & 1.44 & 10.29 & \cellcolor{white}1.00 & 1.00 & 2.42 & \cellcolor{Green!10}76 \\
& A5 & 3.38 & \cellcolor{white}4110 & \cellcolor{white}40.0 & 1.78 & 9.86  & \cellcolor{white}1.00 & 1.00 & 3.38 & \cellcolor{Green!10}88 \\
& A6 & 4.99 & \cellcolor{white}4170 & \cellcolor{white}50.0 & 2.24 & 9.28  & \cellcolor{white}1.00 & 1.00 & 4.99 & \cellcolor{Green!10}99 \\

\caption{Complete list of datasets/cases considered here with corresponding fitting results (see \href{https://cocalc.com/share/public_paths/cd3b9c28f1e14b1de432482bc8009c461a69cfad}{notebooks}). Inputs and outputs of the composite profile fitting method, as described in section~\ref{sec:NickelsLimits}, are highlighted in yellow and green, respectively.}
\label{tab:Alldatasests}
\end{longtable}
\normalsize

\section{Overshoot function}
\label{apx:overshoot}
\begin{figure}
    \captionsetup{width=1.00\linewidth}
    \begin{center}
    \includegraphics[width=0.50\textwidth]{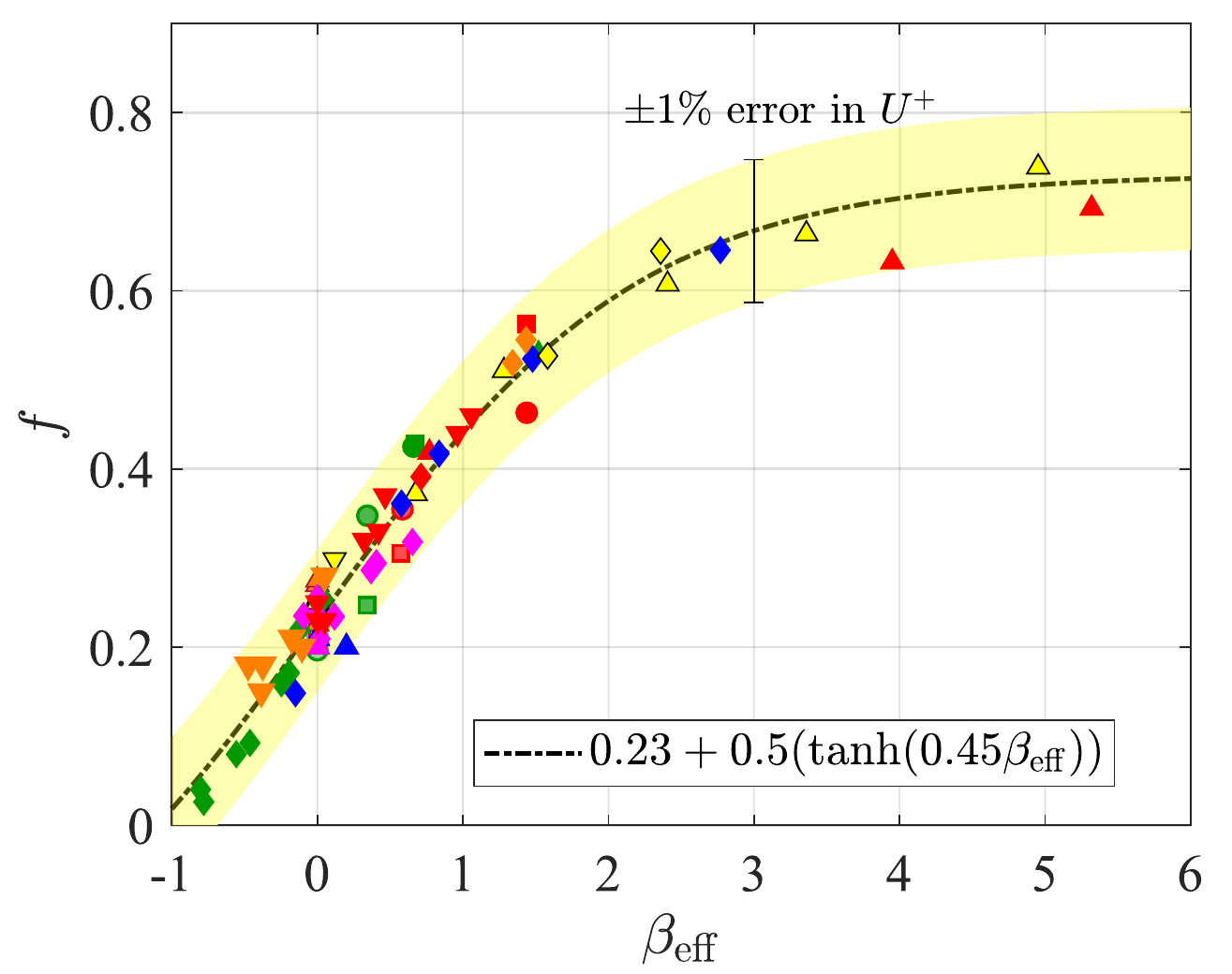}
    \end{center}
    \caption{Maximum overshoot of the mean velocity in the buffer region as a function of $\beta_{\rm eff}$.}
    \label{fig8}
\end{figure}

The maximum overshoot from the log-law in the buffer region, i.e. $f(\zc)$ from \eqref{eq:f}, was measured for all cases in the present study, and the results are plotted as a function of $\beta_{\rm eff}$ \eqref{eq:model} in figure~\ref{fig8}. 
A clear trend is observed: the overshoot increases with $\beta_{\rm eff}$ for the range of APG TBLs considered here; however, the overshoot can not grow unboundedly. 
As such, we propose the following functional form for the maximum overshoot from the log-law as a function of $\zc$, or equivalently $\beta_{\rm eff}$:
\begin{equation}
    f\left(\zc\right) = 0.23 + 0.5 \tanh\left(31 \, \frac{144 - {\zc}^2}{{\zc}^3}\right) = 0.23 + 0.5 \tanh(0.45\beta_{\rm{eff}}).
    \label{eq:fbeta}
\end{equation}
From figure~\ref{fig8}, a clear agreement between this model and the measured overshoots can be observed, with all cases matching within experimental uncertainty. 
Consequently, with the overshoot well-modelled, the proposed composite profile \eqref{eq:PresentInner} can remain a three-parameter composite profile. 
It is also noted that this proposed model \eqref{eq:fbeta} has a formulation consistent with previous studies/datasets, such as \citet{chauhan2009criteria} and \citet{monkewitz2021late}; however, further high-$\re$ studies across a broader range of $\beta_{\rm eff}$ are necessary to refine this formulation. 
The underlying mechanism behind the apparent increase in $f(\zc)$ under an APG can also be analysed through the mean momentum balance (MMB) equation for TBLs \citep{klewicki2013}. 
The MMB equation includes contributions from the mean inertia (MI), pressure gradient (PG), turbulent inertia (TI), and viscous force (VF) terms. 
In the buffer region, the MI term is negligible and, in the absence of any PG, the VF and TI terms are balanced, i.e. ${\partial(VF)}/{\partial(TI)}\approx1$. 
However, by introducing an APG ($dp/dx<0$), this balance shifts, reducing the relative contribution of viscous forces ${\partial(VF)}/{\partial(TI)}<1$ (See figures 1 and 3 from \citealp{romero2022}, and figures 14 and 15 from \citealp{han2024outer}). 
The hypothesis is that this interplay between decreasing viscous and increasing turbulent stresses leads to an imbalance, meaning that Reynolds stress builds up more rapidly than viscous stress can dissipate it, causing the mean velocity profile to overshoot before settling back to the log-law. 

\section{Identifying other upstream history effects}
\label{apx:A}
\begin{figure}
    \captionsetup{width=1.00\linewidth}
    \begin{center}
    \includegraphics[width=1\textwidth]{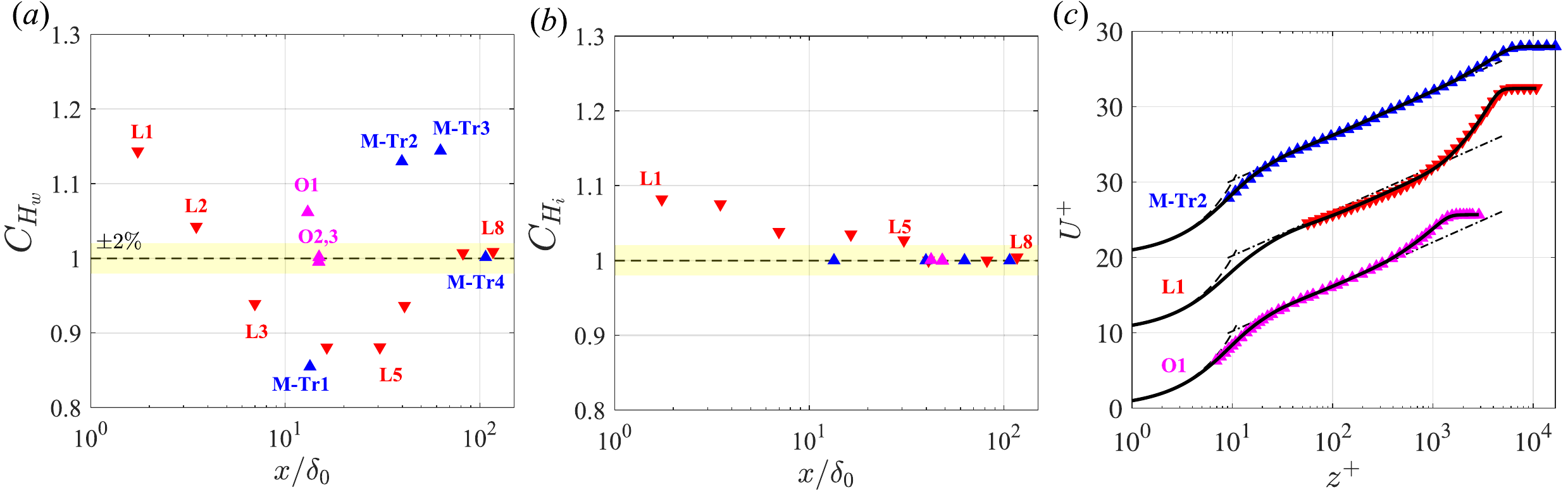}
    \end{center}
    \caption{Variations in \textit{(a)} $C_{H_w}$ and \textit{(b)} $C_{H_i}$, and \textit{(c)} example mean velocity profiles with their corresponding fitted composite profiles for cases with minimal PG history effects but different tripping/surface roughness histories. The shaded yellow region indicates $\pm2\%$ error.}
    \label{Triiping}
\end{figure}

Changes in tripping conditions and/or perturbations in the surface condition/roughness, also leave `history' effects on the TBL mean velocity profile, irrespective of PG history effects, primarily affecting large-scale structures and reshaping the wake region  \citep{marusic2015, orlu2010}. 
Interestingly, we can identify these other history effects using the same composite profile, which can further contribute to the characterisation of `well-behaved' TBLs ($C_{H_i}\approx C_{H_w}\approx 1$). 
Here, we consider datasets from \citet{marusic2015} and \citet{orlu2010}, which employ various tripping conditions, as well as the dataset from \citet{li2021experimental}, which involves a sudden change in surface condition from rough to smooth. 
The calculated values of $C_{H_w}$ and $C_{H_i}$, along with examples of the corresponding composite profiles, are shown in figure~\ref{Triiping}. 
As shown, once the TBL has recovered ZPG-like statistics, consistent with the observations reported in the cited studies, both $C_{H_w}$ and $C_{H_i}$ also converge toward unity. 
In addition, while tripping conditions primarily affect the wake region (changing $C_{H_w}$), it typically leaves the inner/overlap region ($C_{H_i}$) unaffected, as seen in figure~\ref{Triiping}. 
In contrast, for the case with the rough to smooth transition, $C_{H_w}$ and $C_{H_i}$ both change \emph{independently}, and a fundamentally different physical framework must be considered to interpret these changes, e.g. a `blending model' for the mean velocity profile \citep{li2021experimental} where a smooth-wall profile approximates the flow near the wall while the wake region retains rough-wall characteristics. 

\bibliographystyle{jfm}
\bibliography{references}

\end{document}